\documentclass[twocolumn]{aastex701}

\usepackage{graphicx}	
\usepackage{amsmath}	
\usepackage{hyperref} 

\begin{document}

\title{The Impact of Orbital Anisotropy Assumptions in Lensing-Dynamics Modeling}

\author[sname=Liang,gname=Yan]{Yan, Liang}
\affiliation{Department of Astronomy, Tsinghua University, Beijing 100084, China}
\email[show]{liangy19@mails.tsinghua.edu.cn}  

\author[sname=Xu,gname=Dandan]{Dandan, Xu} 
\affiliation{Department of Astronomy, Tsinghua University, Beijing 100084, China}
\email{dandanxu@mail.tsinghua.edu.cn}

\author[orcid=0000-0002-5558-888X, gname=Anowar,sname='J. Shajib']{Anowar J. Shajib}
\altaffiliation{KICP Fellow}
\affiliation{Kavli Institute for Cosmological Physics, University of Chicago, Chicago, IL 60637, USA}
\affiliation{Department of Astronomy and Astrophysics, University of Chicago, Chicago, IL 60637, USA}
\affiliation{Center for Astronomy, Space Science and Astrophysics, Independent University, Bangladesh, Dhaka1229, Bangladesh}
\email{ajshajib@uchicago.edu}

\author[sname=Shu,gname=Yiping]{Yiping Shu}
\affiliation{Purple Mountain Observatory, Chinese Academy of Sciences, Nanjing 210023, China}
\email{yiping.shu@pmo.ac.cn}

\author[sname=Li,gname=Ran]{Ran Li}
\affiliation{School of Physics and Astronomy, Beijing Normal University, Beijing 100875, People’s Republic of China}
\affiliation{Institute for Frontiers in Astronomy and Astrophysics, Beijing Normal University, Beijing 102206, People’s Republic of China}
\email{liran@bnu.edu.cn}


\begin{abstract}

We investigate potential systematic biases introduced by assumptions regarding stellar orbital anisotropy in joint lensing-dynamics modeling. Our study employs the massive early-type galaxies from the TNG100 simulation at redshifts z = 0.2, 0.5, and 0.7. Based on the simulated galaxies, we generate a self-consistent mock dataset containing both lensing and stellar kinematic observables. This is achieved through taking the potential composed of both dark matter and baryons of the simulated galaxies, plus the radial variation of the stellar orbit anisotropy depicted by a logistic function. By integrating constraints from both lensing and stellar kinematics, we separate the contributions of stars and dark matter inside the galaxies. Under three commonly adopted stellar anisotropy assumptions (isotropic orbits, constant anisotropy, and the Osipkov-Merritt profile), the model inferences suggest that the systematic biases in the total stellar mass and central dark matter fraction are not significant. Specifically, the total stellar mass on average is underestimated by less than $0.03\pm0.10$ $\rm dex$ while the dark matter fraction experiences only a statistically insignificant increase of less than $2\%\pm10\%$ at the population level. The dark matter inner density slope in our tests is over-predicted by $0.15\pm0.2$. Additionally, these lacks of significant biases are insensitive to the discrepancies between the assumed anisotropy in modeling and the ground truth orbital anisotropy of mock sample. Our results suggest that conventional assumptions regarding orbital anisotropy, such as an isotropic profile or the Osipkov-Merritt model, would not introduce a significant systematic bias when inferring galaxy mass density distribution at the population level.

\end{abstract}



\section{Introduction} 
According to the $\Lambda$-cold dark matter ($\Lambda$CDM) paradigm, galaxies form within the dark matter halos (\citealt{Rees_Ostriker_1977_GalaxyFormation,White_Frenk_1991_Hierarchical_Clustering,Mo_Mao_White_1998_DiskFormation}) and experience a series of merger or accretion events to assemble its mass and size during evolution (\citealt{Naab_2009_MinorMerger_SizeETG,Oser_2010_GalaxyFormation_2Phase,Hilz_2013_MinorMerger_inside-out_growth,Tortora_2014_EvolutionETG_DM}). The final state of such an evolutionary track is commonly acknowledged as the early-type galaxy (ETG). ETGs observed from the nearby universe to the higher redshifts provide a unique insight into exploring theories of the evolution of galaxies. Specifically, the matter distribution among different components of ETGs is one of the most primary properties entailing the information of its evolutionary history.

In general, the mass content of the ETG consists of two main constitutes: the stars and dark matter halo. For the stellar component, one way to measure stellar mass is stellar population synthesis (SPS) modeling (\citealt{Charlot_Bruzual_1991ApJ_SPS,Bruzual_Charlot_1993_SPS}). However, this method is strongly entangled with the assumption about the mass spectrum of the stars at the birth of the stellar population, known as the stellar initial mass function (IMF). \citet{Salpeter_1955_IMF} summarized a power-law model for the IMF using the main-sequence stars in the solar neighborhood. Subsequently, alternative IMFs with a reduced proportion of low-mass stars were introduced by \citet{Koupa_2001_IMF} and \citet{Chabrier_2003_IMF}. Since the IMF is not only a pivotal assumption in the SPS calculations but also an indicator associated with the environmental conditions during galaxy formation (\citealt{Padoan_Nordlund_2002_IMF_Turbulent_Fragmentation,Bonnell_2006_IMF_JeansMass,Krumholz_2006ApJ_IMF_RadiativeFeedback_MassiveProtoStar,Hennebelle_2008_IMF_CO_Clump&ProtoStarCore,Hopkins_2013_turbulent_fragmentation,Chabrier_2014ApJ_IMF_Starburst}), it is necessary to measure the stellar mass with other independent methods to calibrate the IMF across the galaxy population. For the dark matter, the N-body simulation has outlined that the dark matter halo clusters as a universal mass profile at different mass scales (\citealt{Navarro_1997_NFW}, NFW profile). However, the dark matter mass distribution is also influenced by the baryonic physics and finally deviates those profiles derived from N-body simulations after galaxy formation. For instance, the concentration of the baryons in the halo centre can induce the adiabatic contraction of dark matter (\citealt{Blumenthal_1986_AdiabaticContraction}) while the energy feedback driven by active galactic nucleus (AGN) or star formation will remove the matter from the galaxy centre and flatten the mass profile of the dark matter in the central region (\citealt{Duffy_2010_DM_BaryonicPhysics,Pontzen_Governato_2012_SNfeedback_DM,Governato_2012MNRAS_DM_flattening}).

To detect the mass distribution of ETGs, astronomers have developed a series of observational instruments and theoretical methodologies (reviewed comprehensively by \citet{Courteau_2014_GalaxyMass}). For example, the implementation of the integrated field unit spectroscopy (IFU, \citealt{Bacon_1995_First_IFU}, see \citealt{Cappellari_2016_ETGs_IFU} for a review) enables astronomers to obtain the spatially resolved measurements of the stellar velocity moments along the line-of-sight. These measurements constrain the galaxy mass distributions through the galactic dynamical theory (\citealt{Jeans_1922_JeansModel,Schwarzschild_1979_OrbitModel,Richstone_Tremaine_1984_SphericalGalaxyModel,Binney_Mamon_1982_M2L_Ani_M87,Mamon_Lokas_2005_JeansModel}). However, conducting the IFU-based observations for the galaxies beyond the local universe remains costly. To overcome this limitation, the strong gravitational lensing by massive ETGs can probe the mass distribution of these massive galaxies and investigate the evolution of the ETGs over the cosmic redshifts (reviewed by \citet{Treu_2010_GSLreview,Shajib_2024_GSLreview}). As the lensing model also has its intrinsic degeneracies (e.g., mass-sheet degeneracy, MSD; see \citealt{Gorenstein_1988_MSD,Saha_2000_MST}), the foreground galaxy kinematics such as the stellar velocity dispersion within a single aperture is always combined with the strong lensing modeling (\citealt{Treu_Koopmans_2002_InnerStructure_MG2016112,Treu_Koopmans_2004ApJ_ETG_DMhalo_EvolutionZ<1,Koopmans_2003_StructureDynamics_LensingETG,Koopmans_2006_SLACS3_ETG_StructureFormation,Treu_2010_SLACS_ETG_IMF,Auger_2010_SLACS_HeavierIMF_SL-WL_Dyn,Sonnenfeld_2013_SL2S4_TotalMassProfile,Sonnenfeld_2015_SL2S5_IMF_DMhalo,Sonnenfeld_2018_M2Lgrad_SLACS,Oldham_2018_DM_contraction_&_M2L_gradient,Shajib_2021_Lensing_NFWHalo,Tan_2024_Dinos1,Sheu_2025_Dinos2}). These studies via joint lensing-dynamics modeling collectively provide an overview of the evolution of massive ETGs at $z < 1$. But discrepancies persist among the results. Regarding the stellar IMF, the general trend from both strong lensing and stellar kinematic constraints suggests that galaxies with higher mass or velocity dispersion tend to have a Salpeter-like IMF (\citealt{Smith_2020_M2L_IMF_variation}). However, the exceptions with a Chabrier-like IMF have been identified by \citet{Smith_2015MNRAS_NearbyLens_ChabrierIMF}, \citet{Collier_2018_NewNearbyLens} and \citet{Sonnenfeld_2019_SuGOHI3_ChabrierIMF_Lens}. Meanwhile, for the dark matter halo, the intriguing discrepancies also emerge. On the one hand, the central dark matter mass fraction of the ETGs from simulations (\citealt{XuDandan_2017_Illustris_ETGs,WangYunchong_2020_ETG_TNG1}) is systematically higher than the values reported by strong lensing galaxies (\citealt{Oldham_2018_DM_contraction_&_M2L_gradient}), despite the former having accounted for the effects of baryonic physics and energy feedback (\citealt{Vogelsberger_2014_Illustris,Weinberger_2017_Simulation_EnergyFeedback}). On the other hand, the strong lensing galaxy population itself exhibits a diverse range of possibilities regarding the halo's response to baryons. For example, \citet{Oldham_2018_DM_contraction_&_M2L_gradient} selected 12 early-type/early-type lensing galaxies (\citealt{Oldham_2017_EELS}) and combined lensing image data with stellar kinematics in the inner regions. Their analysis revealed a bimodal distribution in the logarithmic inner density slope of the dark matter, with 4 systems exhibiting an inner slope below 0.5 (the inner slope of the standard NFW profile is 1), while 6 systems showed much steeper, cusped profiles. \citet{Shajib_2021_Lensing_NFWHalo} collected 23 strong lensing systems from SLACS sample (\citealt{Bolton_2006_SLACS1,Auger_2010_SLACS_HeavierIMF_SL-WL_Dyn}) and combined the strong lensing data with weak lensing measurements (\citealt{Gavazzi_2007_WeakLensing_SLACS_ETGs}) and stellar velocity dispersion within the centre. They found that these lensing galaxies have neither significant halo contraction nor expansion compared to the standard NFW model. 

The underlying causes of these discrepancies remain unclear. The baryon physics can provide some possible explanations. For example, the galaxies living in dense environments might undergo more frequent merger or accretion events, which could activate the AGN feedback and flatten the internal dark matter mass profile (\citealt{Oldham_2018_DM_contraction_&_M2L_gradient}).  Furthermore, the limited number of strong lensing galaxies, coupled with the fact that only the lensing systems with more massive galaxies as deflectors possess a sufficiently large cross section and thus are more likely to be detected, introduces the selection bias that may skew statistical results (\citealt{Sonnenfeld_2023_selection_function}). Additionally, systematic biases inherent in the joint lensing-dynamics modeling process need to be quantified to check if they are negligible or not. A significant source of such systematics is the assumption regarding the stellar mass-to-light ratio $M_\star/L$. Given that the stellar mass distribution roughly follows the light distribution, many studies assume that the stellar mass density is directly proportional to light with a uniform $M_\star/L$ within individual galaxies. However, an increasing number of studies have reported the evidence of radial variation in $M_\star/L$ (\citealt{Tortora_2011_M2L_Gradient,Oldham_2018_M87_IMF_radial_variation,LaBarbera_2019_IMF_Gradient_most_massive_ETGs,Smith_2020_M2L_IMF_variation}) \footnote{The spatially uniform stellar mass-to-light ratio is also supported by \citet{Shajib_2021_Lensing_NFWHalo,Sheu_2025_Dinos2}}. The constant $M_\star/L$ assumption for those galaxies with a $M_\star/L$ radial gradient may lead to an underestimation on the dark matter mass in the galaxy centre, even though full constraints on the total mass density distribution with very high spatial resolution are available (\citealt{LiangYan_M2L_bias}). 

Another potential contributor to the systematic biases during the joint lensing-dynamics modeling pertains to the assumption concerning stellar orbit anisotropy. The stellar orbit anisotropy is quantified by the ratio of tangential to radial stellar motions and is a crucial parameter within the framework of the stellar dynamics model (\citealt{Jeans_1922_JeansModel,Binney_Mamon_1982_M2L_Ani_M87,Mamon_Lokas_2005_JeansModel,Binney_Tremaine_2008_Galactic_Dynamics}). Nevertheless, the direct measurement of orbit anisotropy presents an immense challenge for the lensing galaxies, primarily because this quantity is linked to the state of stars in the six-dimensional phase space. In previous works, the astronomers have to incorporate the certain prior assumptions on this quantity into their dynamical modeling. The most straightforward approach involves assuming an isotropic stellar orbital structure (\citealt{Koopmans_2006_SLACS3_ETG_StructureFormation,Treu_2010_SLACS_ETG_IMF,Sonnenfeld_2013_SL2S4_TotalMassProfile,Sonnenfeld_2015_SL2S5_IMF_DMhalo,Oldham_2018_DM_contraction_&_M2L_gradient}). Alternatively, if the distribution function of the dynamical tracers in the phase space can be expressed as a specific combination of the integrals of motions,  the radial profile of stellar orbit anisotropy can be analytically derived, as outlined in the Osipkov-Merritt model (\citealt{Osipkov_1979_OM_model,Merritt_1985_OM_model2,Merritt_1985_OM_model}). This model predicts an isotropic orbital structure in the galaxy centre, monotonically transforming towards the fully radially dominated orbits in the outskirts. It has been adopted by   \citet{Treu_Koopmans_2002_InnerStructure_MG2016112}, \citet{Treu_Koopmans_2004ApJ_ETG_DMhalo_EvolutionZ<1}, \citet{Koopmans_2003_StructureDynamics_LensingETG}, \citet{Sonnenfeld_2012_Jackpot} and \citet{Shajib_2021_Lensing_NFWHalo}. However, both observations and simulations have indicated that the orbit structure of the massive ETG exhibits a nearly isotropic distribution in the innermost regions, changing into some moderately radial anisotropy at larger radii (\citealt{Kronawitter_2000_Orbit_Structure,Onorbe_2007_Mass&VelocityEllipsoid_HydroSim,Koopmans_2009_SLACS_ETGs_Isotropy,Cappellari_2013_ATLAS3D_15,XuDandan_2017_Illustris_ETGs}). 

The purpose of this work is to examine the potential systematic biases arising from the assumptions about stellar orbit anisotropy used in previous research. Specifically, we aim to quantify the impact of prior orbit anisotropy assumptions on the inference of separating stellar and dark matter density profiles of massive ETGs. Besides, we seek to determine which source of systematic bias is predominant, in comparison to the stellar mass-to-light ratio gradient. To accomplish this, we generate the mock lensing and stellar kinematic observables of massive ETGs selected from the hydrodynamical cosmological simulation IllustrisTNG (\citealt{Nelson_TNG_1,Marinacci_TNG_2,Naiman_TNG_3,Pillepich_TNG_4,Springel_TNG_5}). Particularly, the mock stellar kinematic data are crafted to reflect the stellar orbit structures of the simulated galaxy sample. Subsequently, we apply the joint lensing-dynamics model, drawing on the methodology described by \citet{Shajib_2021_Lensing_NFWHalo}, to these simulated galaxies. Our model incorporates a two-component mass distribution (comprising stars and a dark matter halo) and is tested under three commonly adopted assumptions regarding orbit anisotropy: (1) isotropic orbits, (2) constant anisotropy, and (3) the Osipkov-Merritt profile. 

This article is structured as follows: Section~\ref{sec:2} introduces our sample of early-type galaxies from simulations, together with the mock density distributions and mock observable datasets used. Section~\ref{sec:3} describes the methodology of our joint lensing-dynamical modeling framework. The modeling results and their associated systematic biases are reported in Section~\ref{sec:4}. We discuss the inference of the inner dark matter density slope in Section~\ref{sec:5.1}, while Section~\ref{sec:5.2} compares the biases introduced by orbital anisotropy and those due to the stellar mass-to-light ratio gradient investigated in the previous study. In Section~\ref{sec:5.3}, we discuss the impact of stellar orbital anisotropy on the total density slope. Finally, Section~\ref{sec:6} summarizes our main findings and conclusions. The cosmological parameters adopted in this study align with the observation from \citet{Planck_2016_Cosmology}, to maintain consistency with IllustrisTNG simulation, with $\Omega_{\rm m} = 0.3089$, $\Omega_{\Lambda} = 0.6911$ and $H_0 = 67.74 \rm km/s/Mpc$.

\section{Simulated Sample \& Data Set}
\label{sec:2}
\subsection{Massive Early-Type Galaxies in Simulation}
The galaxy sample for this research is extracted from The Next Generation Illustris Simulations (IllustrisTNG, released in 2018; \citealt{Nelson_TNG_1,Marinacci_TNG_2,Naiman_TNG_3,Pillepich_TNG_4,Springel_TNG_5}), which are a suite of state-of-the-art magneto-hydrodynamical cosmological simulations based on the moving mesh code AREPO (\citealt{Springel_AREPO}). The IllustrisTNG dataset includes three simulation volumes: TNG50, TNG100, and TNG300, each characterized by distinct box sizes and resolutions. Specifically, in this study, we chose galaxies of the TNG100 simulation. This simulation employs a box size of $110.7 \rm Mpc$ and a gravitational softening length of $\epsilon_{\rm grav} = 0.74 \rm kpc$. The mass resolutions for baryonic matter and dark matter are $m_{\rm baryon} = 1.4\times10^6 \rm M_\odot$ and $m_{\rm DM} = 7.5\times10^6 \rm M_\odot$, respectively. The galaxy identification is achieved via the SUBFIND algorithm (\citealt{Springel_2001_SUBFIND,Dolag_2009_SUBFIND}).

To investigate biases in joint lensing-dynamics modeling, we focus on massive early-type galaxies at intermediate-to-high redshifts in the simulation. Our sample selection adheres to the criteria outlined in \citet{XuDandan_2017_Illustris_ETGs} and \citet{LiangYan_M2L_bias}. Below, we merely provide a concise summary of the methodology employed. We first select galaxies with stellar masses in the range $ 10.8 < \log{M_\star/\rm M_\odot} < 12.2$ with a high central velocity dispersion ($170\rm km/s < \sigma_{\rm V} < 350 \rm km/s$, luminosity-weighted) at three main snapshots: $z=0.2$, $z=0.5$ and $z=0.7$, which are comparable to the redshifts of the lensing galaxies discovered in SLACS (\citealt{Bolton_2006_SLACS1,ShuYiping_2015_SLACS_XII_Extended_Sample,ShuYiping_2017_SLACS_40NewLens}), SL2S (\citealt{Gavazzi_2012_SL2S_1}) and BELLS (\citealt{Brownstein_2012_BELLS_1}). We then carry out the photometric bulge-disk decomposition for these galaxies. The bulge component is fitted using a de Vaucouleurs profile (\citealt{deVaucouleurs_1948}), whereas the disk is fitted with an exponential profile. Only galaxies with a bulge-to-total light ratio $B/T>0.5$ are classified as ETGs. To ensure robust bulge-dominated morphology, we further fit the light profile with single-component de Vaucouleurs and exponential models, respectively. We retain galaxies for which the de Vaucouleurs model yields a better fit. In addition, we introduce a new criterion for this study. We employed the cored power-law model to fit the galaxy total density profile, keeping those with $1.55< \eta_{\rm tot} < 2.45$ (see Sec.~\ref{sec:5.3} for detailed formulation) in our sample to maintain consistency with the lensing galaxies from SLACS, SL2S and BELLS sample (\citealt{Auger_2010_SLACS10_stellar_total_mass_correlation,Bolton_2012_BELLS2_MassDensityProfile_evolution,Sonnenfeld_2013_SL2S4_TotalMassProfile}). Through this selection procedure, we totally identify 287 massive ETGs across the three snapshots.

\subsection{Mock Data Set}
\label{sec:2.2}
With our selected massive early-type galaxy sample, we generate mock observables of gravitational lensing and stellar kinematics to perform subsequent joint lensing-dynamical modeling. Following the methodology and observational dataset presented by \citet{Sonnenfeld_2018_M2Lgrad_SLACS} and \citet{Shajib_2021_Lensing_NFWHalo}, we adopt four types of observable for each galaxy: (1) Einstein radius $\theta_{\rm E}$ which reflects the total projected mass within it; (2) dimensionless radial quantity $\xi = \theta_{\rm E}\alpha^{\prime\prime}(\theta_{\rm E})/(1-\kappa(\theta_{\rm E}))$, where $\alpha^{\prime\prime}(\theta)$ is the second derivative of deflection angle and $\kappa(\theta_{\rm E})$ is the convergence at $\theta_{\rm E}$, this quantity is the MSD-invariant "reduced slope" that the lensing imaging data constrain (\citealt{Kochanek_2020_Xi,Birrer_2021_LensingFormalism_CurvedArcBasis}); (3) the tangential shears from the weak lensing signals $\gamma_{\rm t}$, probing the outer mass profile beyond the Einstein radius; (4) the line-of-sight stellar velocity dispersion $\sigma_{\rm los,e/2}$ averaged within $0.5R_{\rm eff}$, which traces the central gravitational potential. Here, $R_{\rm eff}$ is the effective radius of the simulated galaxy, defined as the radius of a sphere enclosing half of the total stellar mass. In this subsection, we will provide a thorough explanation of the construction of our mock datasets. Specifically, the definitions of $\theta_{\rm E}$, $\xi$, and $\gamma_{\rm t}$ are explained in the Sec.~\ref{sec:2.2.2}. The mock stellar kinematics data generation are introduced in Sec.~\ref{sec:2.2.3}.

\subsubsection{Mock Mass Density Distribution}
\label{sec:2.2.1}
The most straightforward approach for mocking lensing or kinematics data is to extract the relevant information from the particle data of these simulated galaxies. For instance, the central velocity dispersion of a simulated galaxy can be derived as either mass- or luminosity-weighted standard deviation of velocity along the line-of-sight of all stellar particles within the given aperture. However, two main factors complicate this process. On the one hand, the kinematics of a simulated galaxy are associated with not only its morphology but also kinematic substructures formed from the infalling cold gas. The latter is not commonly accounted for in the mass distribution adopted in dynamics model. On the other hand, numerical effects, such as gravitational softening or spurious heating of stellar motions, can introduce additional biases into the stellar kinematics (\citealt{Ludlow_2019_EnergyRepartition_Simulation,Ludlow_2023_SpuriousHeating_Simulation}), particularly in the central region where the stellar kinematics concerned in joint lensing-dynamics modeling. Thus, we discard the direct simulation output as the mock observation, but rather reconstruct consistent galaxy mass density and stellar kinematic distribution with the method described below. We first apply the analytical multi-Gaussian expansion method (MGE, \citealt{Emsellem_1994_MGE}) to represent the mass density distribution of both the stars and the dark matter halo of our sample galaxies. Next, we use this mock mass density distribution to generate the mock stellar kinematic data via the spherical Jeans modeling (\citealt{Jeans_1922_JeansModel}). This process enables us to maintain consistency between the mock stellar kinematic data and our following lensing-dynamics model. Here, the formulation is shown below:
\begin{equation}
    \rho^{(\rm mock)}_{\rm tot} = \rho^{(\rm mock)}_{\rm st}+\rho^{(\rm mock)}_{\rm dm}, 
\end{equation}
\begin{equation}
\label{eq:MGE_st}
    \rho^{(\rm mock)}_{\rm st}(r) = \sum^{N_{\rm st}}_{\rm i=1}\frac{M_{\rm st,i}}{(2\pi\sigma_{\rm st,i})^{\frac{3}{2}}}\exp{(-\frac{r^2}{2\sigma_{\rm st,i}^2})},
\end{equation}
\begin{equation}
\label{eq:MGE_dm}
    \rho^{(\rm mock)}_{\rm dm}(r) = \sum^{N_{\rm dm}}_{\rm i=1}\frac{M_{\rm dm,i}}{(2\pi\sigma_{\rm dm,i})^{\frac{3}{2}}}\exp{(-\frac{r^2}{2\sigma_{\rm dm,i}^2})},
\end{equation}
where $\rho^{(\rm mock)}_{\rm tot}$, $\rho^{(\rm mock)}_{\rm st}$ and $\rho^{(\rm mock)}_{\rm dm}$ are $3D$ analytical mock total, stellar and dark matter mass density distribution, respectively. And $r$ denotes the spherical distance from the galaxy center. The mock stellar (dark matter) mass density profile is represented as the sum of $N_{\rm st}$ ($N_{\rm dm}$) Gaussian components. Each Gaussian component is characterized by its total mass $M_{\rm st,i}$ ($M_{\rm dm,i}$) and width $\sigma_{\rm st,i}$ ($\sigma_{\rm dm,i}$). We apply such a MGE-based density model to fit both the stellar (within $7R_{\rm eff}$) and the dark matter (within the virial radius) density profiles, which are directly extracted from the particle data of these simulated galaxies. The fittings are implemented using \texttt{MgeFit}\footnote{\texttt{MgeFit} v6.0.4, \href{https://pypi.org/project/mgefit/}{https://pypi.org/project/mgefit}}, a Python routine developed by \citet{Cappellari_2002_MGE}. All subsequent procedures for generating mock observable data are conducted relying on this analytical mass density distribution expressed in the MGE form. 

It is noteworthy that the luminosity distribution of the mock galaxies $l_{\rm st}^{\rm (mock)}(r)$ should also be expressed in the MGE form for the subsequent discussions. As mentioned in the introduction, the luminosity of the galaxy is the constraint to refine the stellar component in the lensing-dynamics modeling. This approach may introduce extra biases in our inference since the true stellar mass-to-light ratio in the mock system $\rho^{(\rm mock)}_{\rm st}(r)/l_{\rm st}^{\rm (mock)}(r)$ is not spatially uniform. To isolate the effect of the orbital anisotropy, we fix the mock stellar mass-to-light ratio at $\rho^{(\rm mock)}_{\rm st}(r)/l_{\rm st}^{\rm (mock)}(r) = 1 \rm M_\odot/L_{\rm \odot}$, equivalently replacing the mock luminosity with the mock stellar mass density. We also use this mock luminosity distribution to generate the mock stellar kinematics in Sec.~\ref{sec:2.2.3}.

Although the spherical symmetry adopted here is consistent with the following lensing-dynamics model, it is not well-suited for galaxies with highly ellipsoidal stellar morphologies\footnote{The dark matter halo of the simulated galaxy is generally spheroidal in the central region. Within a radius of $30 \rm kpc$, the axial ratios of the halo are $\langle d_{\rm b}/d_{\rm a} \rangle = 0.97\pm 0.02$ and $\langle d_{\rm c}/d_{\rm a} \rangle = 0.76\pm 0.07$, where $d_{\rm a}>d_{\rm b}>d_{\rm c}$ are the length of semi-major axes of inertia ellipsoid. More details can be found in \citealt{Allgood_2006_DM_Halo_Shape,XuDandan_2017_Illustris_ETGs}.}. Such morphological mismatches induce systematic errors particularly during the density projection process for producing lensing and stellar kinematic observables. For instance, assuming spherical symmetry may cause systematic biases of a few percent in the measured lensing time-delay. (\citealt{HuangXiangyu_2025_TDCOSMO21_triaxiality&projection}). Since this work focuses on the effect of anisotropy parametrization. So we restrict our analysis to ETGs that are not significantly affected by non-spherical morphology during projection. Specifically, we require that the best-fit MGE profile of the stellar component can not only faithfully capture the $3D$ stellar mass density distribution, but also reproduce the $2D$ surface mass density of the stars after projection. To enforce this, we adopt the profile goodness-of-fit metric from \citet{LiangYan_M2L_bias} to select ETGs with a reliable MGE decomposition:
\begin{equation}
    \langle\Delta\log\rho_{\rm st}\rangle_{\rm j} = \frac{1}{N_{\rm \rho , j}}\sum_{r_{\rm i}<R_{\rm ap,j}}|\log\rho^{\rm (mock)}_{\rm st,i}-\log\rho^{\rm (sim)}_{\rm st,i}|,
\end{equation}
\begin{equation}
    \langle\Delta\log S_{\rm st}\rangle_{\rm j} = \frac{1}{N_{\rm S, j}}\sum_{R_{\rm i}<R_{\rm ap,j}}|\log S^{\rm (mock)}_{\rm st,i}-\log S^{\rm (sim)}_{\rm st,i}|.
\end{equation}
In the metrics above, $\rho^{\rm (mock)}_{\rm st,i}$ and $\rho^{\rm (sim)}_{\rm st,i}$ are the stellar mass density in $\rm i_{\rm th}$ radial bin from the mock MGE profile and the simulation, respectively. And the corresponding surface mass density are denoted by $S^{\rm (mock)}_{\rm st,i}$ and $S^{\rm (sim)}_{\rm st,i}$. For both $\langle\Delta\log\rho_{\rm st}\rangle$ and $\langle\Delta\log S_{\rm st}\rangle$, we evaluate the metrics within four apertures $R_{\rm ap, j} = \{0.2R_{\rm eff},0.5R_{\rm eff},R_{\rm eff},7R_{\rm eff}\}$. We restrict our analysis on the massive ETGs whose all $\langle\Delta\log\rho_{\rm st}\rangle_{\rm j}$ and $\langle\Delta\log S_{\rm st}\rangle_{\rm j}$ values remain below 0.05 $\rm dex$. Following this filtering process, our sample finally retains 146 ETGs. 

\begin{figure}
    \centering
    \includegraphics[width=\linewidth]{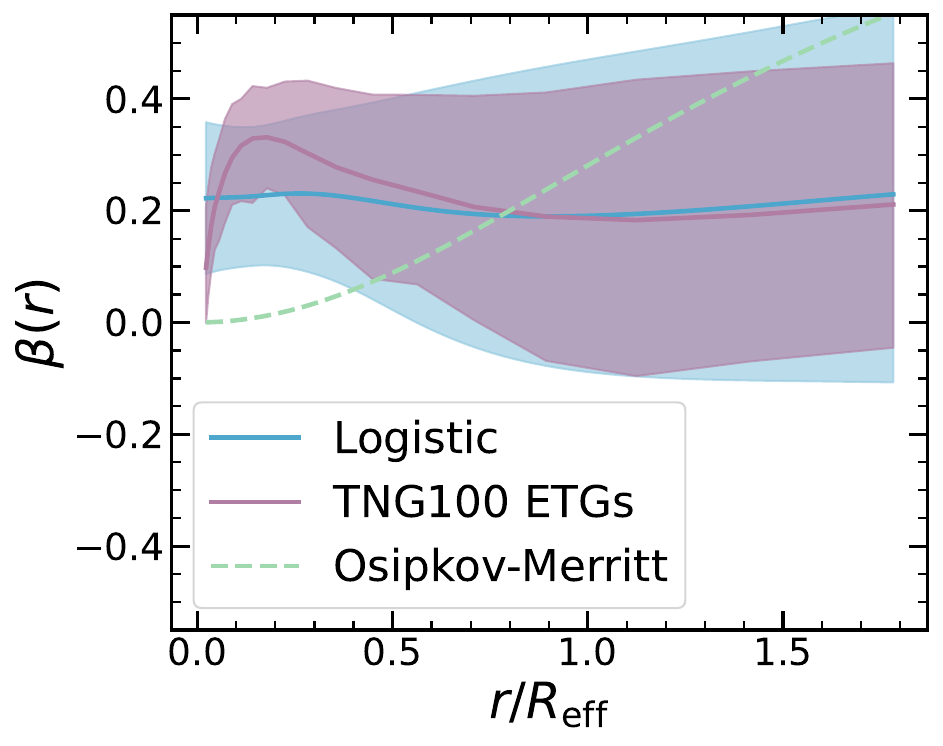}
    \caption{The stacked stellar orbit anisotropy profile of the massive early-type galaxies in TNG100. The purple curve shows the mean anisotropy profile, with the shaded region indicating the $1\sigma$ level scatter across the sample. For comparison, the dashed curve in light green is the Osipkov-Merritt model with a transition radius of $1.6R_{\rm eff}$ (\citealt{Shajib_2021_Lensing_NFWHalo}). The stacked mock anisotropy and its $1\sigma$ scatter from the logistic fitting (see Sec.~\ref{sec:2.2.3}) are depicted by the blue solid curve and the shaded area, respectively.}
    \label{fig:tng_anisotropy}
\end{figure}

\begin{figure*}
    \centering
    \includegraphics[width=\textwidth]{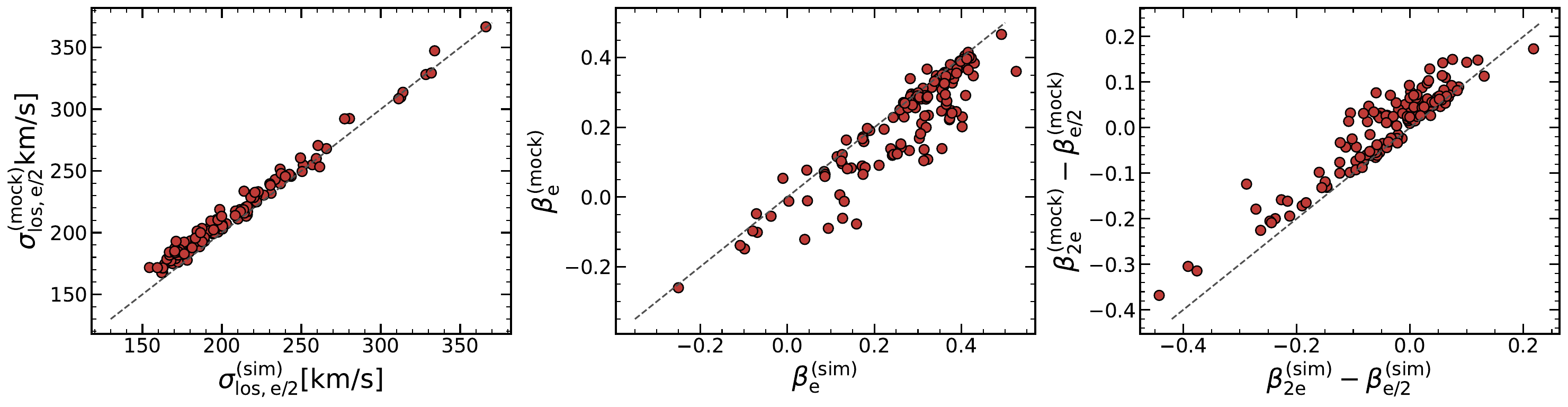}
    \caption{The comparison of the mock stellar kinematics synthesized from smoothed mass profiles and that directly obtained from the motion of stellar particles. \textit{Left:} velocity dispersion integrated along the line-of-sight within $0.5R_{\rm eff}$. Each data point represents one individual simulated galaxy. The y- and x-axis are the results extracted from the kinematics synthesized from mock mass profiles and those directly obtained from the motion of stellar particles, respectively. The dashed line is $y=x$. \textit{Mid:} the comparison of averaged orbital anisotropy within effective radius $R_{\rm eff}$. \textit{Right:} the comparison of the radial gradient of orbital anisotropy defined as the difference between its averaged value within $2R_{\rm eff}$ and $0.5R_{\rm eff}$. The detailed formulations for these averaged anisotropies, such as $\beta_{\rm e}^{\rm (mock)}$ can be found in Appendix \ref{sec:mock_anisotropy}. }
    \label{fig:mock_kinematics}
\end{figure*}

\subsubsection{Mock Lensing Observables}
\label{sec:2.2.2}
Having established the mock density profile in multi-Gaussian expansion form, we now use it to generate two sets of mock observables: observables from imaging data of a given lens system and stellar velocity dispersion. Here, we first detail the procedures for mocking lensing observables.
The first observable is the Einstein radius, which is determined by the enclosed projected mass of the foreground galaxy via:
\begin{equation}
\label{eq:theta_E}
    \theta_{\rm E} = \sqrt{\frac{D_{\rm ls}}{D_{\rm l}D_{\rm s}}\frac{4GM(\theta_{\rm E})}{c^2}},
\end{equation}
where $D_{\rm ls}$ denotes the angular diameter distance between lens and source, $D_{\rm l}$ and $D_{\rm s}$ are the angular diameter distances from observer to lens and source, respectively. The term $M(\theta_{\rm E})$ is the projected mass enclosed within $\theta_{\rm E}$. Here, $G$ represents the gravitational constant and $c$ is the speed of light. We utilize the MGE-based mock mass density (see Eq.~\ref{eq:MGE_st} and Eq.~\ref{eq:MGE_dm}) to represent the projected enclosed mass. The redshift of the background source $z_{\rm s}$ is located at 0.5, 1.5 or 2.4 to keep the Einstein radius roughly $1\rm arcesc$ for the isothermal system with a velocity dispersion of $250\rm km/s$, corresponding to the foreground lens at $z_{\rm l}$ = 0.2, 0.5 or 0.7, respectively. The mock Einstein radius is then obtained by numerically solving Eq.~\ref{eq:theta_E}. 

The other mock observable associated with the strong lensing effect is the dimensionless reduced slope $\xi = \theta_{\rm E}\alpha^{\prime\prime}(\theta_{\rm E})/(1-\kappa(\theta_{\rm E}))$, which is a model-independent parameter introduced by \citet{Kochanek_2020_Xi,Kochanek_2021_Xi}. It can serve as a probe for constraining the mass density slope ($\xi = 2(\eta_{\rm tot}-2)$ for a $3D$ total mass density $\rho_{\rm tot}(r) \propto r^{-\eta_{\rm tot}}$). This quantity is also proportional to the normalized differential radial magnification $\partial_{\rm r}\mu_{\rm r}/\mu_{\rm r}$ at $\theta_{\rm E}$. Therefore, it can be measured through the relative radial thickness of distinct arcs or the positions of an image pair of the same source relative to the critical curve (further details and formulations can also be found in \citealt{Sonnenfeld_2018_Xi,Birrer_2021_LensingFormalism_CurvedArcBasis,Shajib_2024_GSLreview}). Adopting the similar approach as described above, we employ the MGE-based analytical mass density distribution to generate both the convergence $\kappa(\theta)$ and the deflection angle $\alpha(\theta)$ of the mock lensing system. Once these quantities are obtained, the parameter $\xi$ can be easily derived. 

In order to access the constraining power on the mass distribution at large radii, we also prepare the mock data to simulate the weak lensing measurements. Generally, the weak lensing effects are predominantly detected around extremely massive objects such as galaxy clusters. A few measurements and analyses have been conducted on the galaxy scale (\citealt{Gavazzi_2007_WeakLensing_SLACS_ETGs,Auger_2010_SLACS_HeavierIMF_SL-WL_Dyn,Sonnenfeld_2018_M2Lgrad_SLACS,Shajib_2021_Lensing_NFWHalo}). In this research, we use the tangential shear, which is proportional to the excess projected mass density $\gamma_{\rm t}\Sigma_{\rm crit} = \langle\Sigma(<R)\rangle-\Sigma(R)$ ($\Sigma_{\rm crit}$ is the critical surface density of the lensing system, see \citealt{Schneider_2005_WeakLensingReview} for details), as our mock constraints. Here, $\Sigma(R)$ represents the surface density of the lens at projected radius $R$ from the galaxy centre while $\langle\Sigma(<R)\rangle$ is the averaged surface density within $R$, both of which can be straightforwardly derived from our mock analytical mass distribution. Specifically, we follow the same methodology presented by \citet{Shajib_2021_Lensing_NFWHalo}, taking $\gamma_{\rm t}$ within four logarithmically spaced radial bins spanning 10-100 kpc to constrain the outer mass distribution of our sample. It should be noted that the real observation of weak lensing effects relies on the multiple background sources to provide the statistical signals. However, increasing the number of sources only enhances the signal-to-noise ratio of the measurements. We can directly assign an uncertainty of observational level to the tangential shear to mimic the statistical effect of multiple sources. Consequently, for the purpose of mocking data for the following modeling, the inclusion of multiple source planes is unnecessary.   

We emphasize that our Bayesian framework for joint lensing-dynamical modeling incorporates the error propagation of the observational level. To be specific, an uncertainty of $0.01 \rm arcsec$ is assigned to the mock $\theta_{\rm E}$ measurement (\citealt{LiTian_2024_GGSLpopulation,Sonnenfeld_2025_debiased_SLACS_sample2}). The radial parameter $\xi$ retains the uncertainty of 0.2, roughly consistent with \citet{Shajib_2022_TDCOSMO_IX}. The weak lensing shear (or equivalently, excess surface density) is modeled with a conservative 30\% uncertainty (\citealt{Gavazzi_2007_WeakLensing_SLACS_ETGs}).

\subsubsection{Mock Stellar Kinematics}
\label{sec:2.2.3}
The stellar kinematics within the central region of the galaxy serve as another important constraint, aiding in breaking the mass-sheet degeneracy in the lensing model. As previously stated, we generate
the mock stellar kinematics through the spherical Jeans modeling. In this procedure, the mock stellar kinematic data consist of the line-of-sight central velocity dispersion within $0.5R_{\rm eff}$ and the stellar orbital anisotropy. We expect that the former aligns with the line-of-sight velocity dispersion profiles directly obtained from the stellar particle motions, whereas the latter is roughly consistent with the velocity dispersion ellipsoid of the simulated galaxy at the population level.

Following the procedure for mocking lensing observables, we employ the analytical MGE-based density distribution in the spherical Jeans model:
\begin{equation}
\label{eq:Jeans}
    \frac{{\rm d}l(r)\overline{v^2_{\rm r}}(r)}{{\rm d}r}+\frac{2\beta}{r}l(r)\overline{v^2_{\rm r}}(r) = -l(r)\frac{GM(r)}{r^2},
\end{equation}
where $M(r)$ is the total mass spherically enclosed within $r$ while $l(r)$ is the $3D$ stellar luminosity at $r$ (\citealt{Jeans_1922_JeansModel,Mamon_Lokas_2005_JeansModel,Binney_Tremaine_2008_Galactic_Dynamics}). $\overline{v^2_{\rm r}}(r)$ is the second moment of the radial velocity. Here we note that the first moments of velocity naturally vanish under spherical symmetry, so we have $\overline{v^2_{\rm r}}(r) =\sigma_{\rm r}^2$ and $\overline{v^2_{\rm t}}(r) =\sigma_{\rm t}^2$. The stellar orbital anisotropy parameter $\beta$ (anisotropy, hereafter) is defined by the ratio of the tangential velocity dispersion $\sigma_{\rm t}^2 = (\sigma^2_{\theta}+\sigma^2_{\phi})/2$ and the radial velocity dispersion $\sigma_{\rm r}^2$ via:
\begin{equation}
    \beta = 1-\frac{\sigma_{\rm t}^2}{\sigma_{\rm r}^2} = 1-\frac{\overline{v^2_{\rm t}}(r)}{\overline{v^2_{\rm r}}(r)}.
\end{equation}
According to this definition, $\beta = 0$ corresponds to isotropic orbits. When $\beta = 1$, the stellar orbits are purely radial. In the limit $\beta \to -\infty$, the system becomes entirely tangentially dominated, approaching circular orbits. Solving Eq.~\ref{eq:Jeans}, the velocity dispersion projected along the line-of-sight $\sigma^2_{\rm los}(R)$ can be expressed as:
\begin{equation}
\label{eq:Jeans_solution}
    \sigma^2_{\rm los}(R) = \frac{2G}{I(R)}\int_R^\infty K_{\beta}(u)\frac{l(r)M(r)}{r}{\rm d}r,
\end{equation}
where the variable $u$ represents the ratio of the projected radius $R$ to the spherical radial distance $r$. The surface brightness profile is denoted by $I(R)$. The form of the auxiliary kernel function $K_\beta(u)$ depends on the anisotropy $\beta$ (see the appendix of \citealt{Mamon_Lokas_2005_JeansModel} for more details). Here, we utilize the mock density distribution to the Eq.~\ref{eq:Jeans_solution} to express the total enclosed mass $M(r) = \int_0^r 4\pi {r^\prime}^2 \rho_{\rm tot}^{\rm (mock)}(r^\prime){\rm d}r^\prime$. 
Accordingly, the stellar luminosity $l(r)$ and the corresponding surface brightness $I(R)$ are replaced by $l_{\rm st}^{\rm (mock)}(r)$ and its projection, respectively. To numerically compute the integral in Eq.~\ref{eq:Jeans_solution}, we use the MGE-based python software \texttt{JamPy}\footnote{\texttt{JamPy} v7.1.0, https://pypi.org/project/jampy} which is robust Jeans equation solvers designed by \citet{Cappellari_2008_JAM_cylinderical,Cappellari_2020_JAM_spherical}. 

A key unresolved question to mocking the stellar kinematics is how to consider the anisotropy and the appropriate functional form of $K_\beta(u)$. In Fig.~\ref{fig:tng_anisotropy}, we show the stacked anisotropy profile of our simulated ETG sample. The anisotropy is nearly isotropic in the innermost region ($<0.1R_{\rm eff}$) and dramatically increases to the mildly radial-dominated ($\beta \sim 0.3$) within $0.5R_{\rm eff}$. However, the large scatter implies the substantial galaxy-to-galaxy variation in the anisotropy. To better capture this diversity in anisotropy and generate the mock stellar kinematics with the anisotropy consistent with the simulated sample at the population level, we adopt a more flexible anisotropy model: the logistic profile for $\beta(r)$ (\citealt{Baes&vanHese_2007_LogisticAnisotropy}), which can be expressed as
\begin{equation}
\label{eq:logistic}
    \beta^{\rm (mock)}(r) = \beta_0+\frac{\beta_\infty-\beta_0}{1+(r_{\rm a}/r)^a},
\end{equation}
where $\beta_0$ and $\beta_\infty$ denote the asymptotic values of $\beta(r)$ at the galactic center and at large radii. Now, the only parameters to be determined in the right side of Eq.~\ref{eq:Jeans_solution} are $\beta_0$, $\beta_\infty$, $r_{\rm a}$ and $a$. We fit both velocity dispersion profile along the line-of-sight $\sigma_{\rm los}^{\rm (sim)}(R)$ (with the solution of Jeans equation Eq.~\ref{eq:Jeans_solution} using the mock mass density distribution and logistic anisotropy model Eq.~\ref{eq:logistic}) and the anisotropy profile $\beta^{\rm (sim)}(r)$ (with the logistic anisotropy model Eq.~\ref{eq:logistic}). Both $\sigma_{\rm los}^{\rm (sim)}(R)$ and $\beta^{\rm (sim)}(r)$ are derived from the stellar particle motions of our simulated galaxies (see Appendix~\ref{sec:mock_anisotropy} for details). After fitting, the parameters that define the logistic anisotropy profile will be determined. The value of mock central velocity dispersion $\sigma^{\rm (mock)}_{\rm los,e/2}$ is then given by the Jeans equation with the best-fit $\beta_0$, $\beta_\infty$, $r_{\rm a}$ and $a$. Eventually, we generate a mock dynamical system. This system has a mass density distribution represented by a multi-Gaussian expansion series, fitted to the given simulated galaxy, thus making the lensing and dynamical observables self-consistent. Additionally, its line-of-sight velocity dispersion $\sigma^{\rm (\rm mock)}_{\rm los,e/2}$ will be consistent with that of the simulated sample galaxy $\sigma^{\rm (\rm sim)}_{\rm los,e/2}$, thereby its anisotropy $\beta^{\rm (mock)}(r)$ also resembles the orbital structure $\beta^{\rm (sim)}(r)$ of the simulated sample. Therefore, the mock line-of-sight velocity dispersion depends on the mass profile of the given simulated galaxy, with the anisotropy profile following the sample mean of the simulated galaxies.

\begin{figure*}
    \centering
    \includegraphics[width=\textwidth]{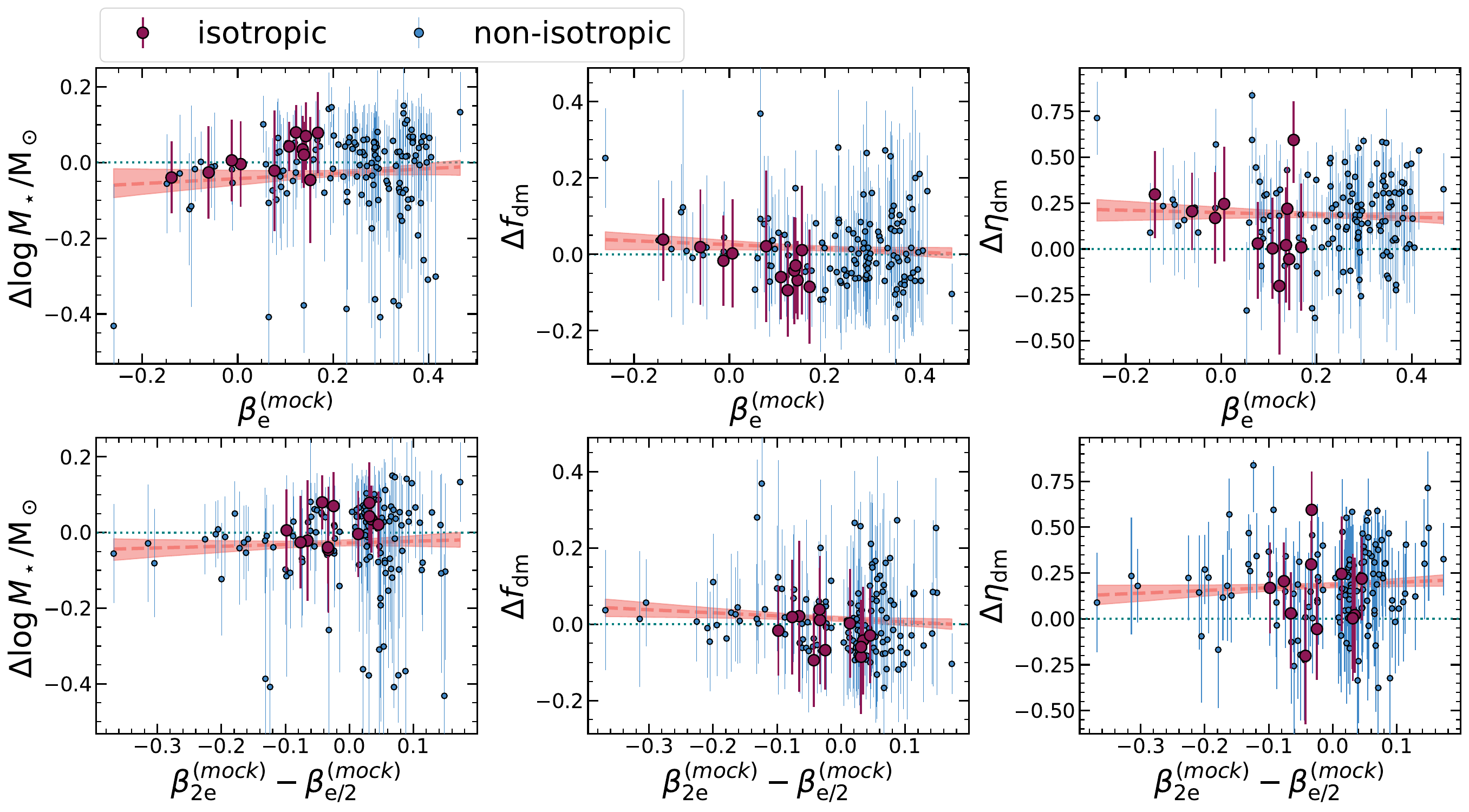}
    \caption{The biases of model parameters under the isotropic orbit assumption. The y-axes display the discrepancies between parameter values estimated by our joint lensing-dynamics model and their true values, so the horizontal dotted line indicates the unbiased results. The quantities appearing in the x-axes $\beta^{(\rm mock)}_{\rm 2e}$, $\beta^{(\rm mock)}_{\rm e}$, $\beta^{(\rm mock)}_{\rm e/2}$ are the averaged anisotropy within $2R_{\rm eff}, R_{\rm eff}, 0.5R_{\rm eff}$ of our mock stellar kinematics (see Appendix~\ref{sec:mock_anisotropy}). The purple points denote galaxies with nearly isotropic orbits, while the blue points represent anisotropic systems. Here, the isotropic and anisotropic systems are classified according to the mock logistic anisotropy profile (see Sec.~\ref{sec:4.1}). The dark red dashed line is the best-fit linear function depicting the dependencies of the systematic biases on the anisotropy (top panels) as well as its radial gradient (bottom panels). The shaded area shows the $1\sigma$ error of this linear fit.}
    \label{fig:results_iso}
\end{figure*}

In Fig.~\ref{fig:mock_kinematics}, we present the comparison between the mock stellar kinematics and the results derived from the stellar particle motions. It is obvious that our mock kinematics can accurately represent the velocity dispersions integrated along the line-of-sight within an aperture radius of $0.5R_{\rm eff}$. Regarding the anisotropy, the middle panel displays the averaged anisotropy within $R_{\rm eff}$, while the right panel illustrates its radial gradient. As shown in these panels, the mock anisotropy is consistent with that of the original simulated galaxy sample at the population level. It is worthy to note that due to the steep transition of the logistic function toward the value of $\beta_0$, the averaged $\beta$ will be naturally smaller than the value from the simulation. Additionally, the anisotropy of four galaxies cannot be successfully represented by our mock kinematics, they are subsequently removed from the sample. 

Building on the above discussion of our mock kinematics, we now quantify the uncertainty associated with our last mock observable: the central velocity dispersion $\sigma^{\rm (\rm mock)}_{\rm los,e/2}$. We adopt an uncertainty of $3.5\%$ for this quantity, consistent with the estimation for the SDSS data by \citet{Knabel_2025_TDCOSMO_19_subpsercent_measurement_kinematics}. 

\begin{table}
    \centering
    \caption{The range of the prior distribution for model parameter in vector $\boldsymbol{
    \omega}$}
    \begin{tabular}{cccccc}
        \hline
        \hline
              &  $\log{M_\star/\rm M_\odot}$ & $f_{\rm dm}$ & $\eta_{\rm dm}$ & $\beta^{\rm (mod)}$ & $\log{r_{\rm om}/R_{\rm eff}}$\\
        \hline
        upper & 13.0 & 0.99 & 1.8 & 1.0  & 0.7 \\
        lower &  9.0 & 0.01 & 0.2 & -0.5 & -1.0 \\
        \hline
    \end{tabular}
    \label{tab:prior}
\end{table}

\section{Joint Lensing-Dynamics Model}
\label{sec:3}
\subsection{Bayesian Framework}
In the joint lensing-dynamics modeling, we combine the mock lensing observables with measurements of the central stellar velocity dispersion to build our dataset. All of these mock observables can be predicted using a mass model of the galaxy. Conversely, the parameters that characterize the mass model can be determined through Bayesian inference. In this subsection, we will provide a detailed account of the entire Bayesian inference workflow.

The mass model of the sample galaxy consists of two primary components: the stars and the dark matter halo. As previously stated, the stellar mass density in lensing-dynamics model $\rho_{\rm st}^{\rm (mod)}$ is scaled by the light of the mock system, whose profile is kept fixed for a given galaxy. The total stellar mass $\log{M_\star/\rm M_\odot}$ is the only free parameter contributed by the stellar component:
\begin{equation}
\label{eq:m2l}
    \rho^{(\rm mod)}_{\rm st}(r) = \frac{M_\star l_{\rm st}^{\rm (mock)}(r)}{\int_0^\infty 4\pi r^{2}l_{\rm st}^{\rm (mock)}(r) {\rm d}r},
\end{equation}
where $l_{\rm st}^{\rm (mock)}$ is the mock luminosity distribution expressed in the MGE form and it is scaled as $\rho_{\rm st}^{\rm (mock)}(r) \rm L_\odot/M_\odot$ (see Sec.~\ref{sec:2.2.1}). To examine the robustness of our findings, we also implement a lensing-dynamical model in which the stellar component is described by a double-S{\'e}rsic profile. In the main context, we only present the results using the MGE-form light distribution. The comparison with results based on the double-S{\'e}rsic light profile is provided in Appendix~\ref{sec:double_sersic}. Overall, the results obtained from the different stellar mass models show good agreement.

On the dark matter side, we adopt the parametric generalized NFW model (\citealt{Zhao_1996_gNFW,Wyithe_2001_gNFW_Lens}), which is given as
\begin{equation}
    \rho_{\rm dm}^{\rm (mod)}(r) = \frac{\rho_{\rm dm,0}}{(\frac{r}{r_{\rm dm}})^{\eta_{\rm dm}}(1+\frac{r}{r_{\rm dm}})^{3-\eta_{\rm dm}}},
\end{equation}
where $\eta_{\rm dm}$ is the inner slope that modifies the compactness of the dark matter halo in the galaxy centre, $\rho_{\rm dm,0}$ is scaled with the dark matter fraction $f_{\rm dm}$ enclosed within a spherical volume of $R_{\rm eff}$. The scale radius $r_{\rm dm}$ is fixed at $r_{\rm dm} = 10R_{\rm eff}$ (\citealt{Sonnenfeld_2015_SL2S5_IMF_DMhalo}).

Upon the mass model, the vector of the parameter space $\boldsymbol{\omega}$ primarily includes the total stellar mass $\log{M_\star/\rm M_\odot}$, the central dark matter fraction $f_{\rm dm}$ and the inner density slope of the dark matter profile $\eta_{\rm dm}$. According to the Bayesian theorem, the posterior probability distribution of the parameter vector $\boldsymbol{\omega
}$ can be given by
\begin{equation}
    P(\boldsymbol{\omega}|\boldsymbol{d}) \propto P(\boldsymbol{d}|\boldsymbol{\omega})P(\omega),
\end{equation}
where $\boldsymbol{d} =\{\theta_{\rm E},\xi,\gamma_{\rm t},\sigma_{\rm los,e/2}\}$ is the vector of mock observed data. $P(\boldsymbol{\omega)}$ is the prior distribution of the model parameter. We adopt a flat prior probability distribution for all parameters. The detailed range for each parameter is listed in Tab.~\ref{tab:prior}. The likelihood function $P(\boldsymbol{d}|\boldsymbol{\omega})P(\omega)$ is decomposed into a product of individual likelihood terms:
\begin{equation}
    P(\boldsymbol{d}|\boldsymbol{\omega}) = P(\theta_{\rm E}|\boldsymbol{\omega}) P(\xi|\boldsymbol{\omega}) P(\sigma_{\rm los,e/2}|\boldsymbol{\omega}) \prod_{i} P(\gamma_{{\rm t},i}|\boldsymbol{\omega}).
\end{equation}

In this product formulation, each component on the right-hand side corresponds to a Gaussian distribution of a mock observable, with the mean given by the corresponding lensing or spherical Jeans model, following the similar procedure with the mock data generation process. The difference in this step is that we employ the aforementioned mass model: stellar mass density scaled by the mock luminosity and gNFW model parametrized by $\boldsymbol{\omega}$. And the standard deviation of each Gaussian likelihood term is aligned with the corresponding observational uncertainties (see the end of Sec.~\ref{sec:2.2.2} and Sec.~\ref{sec:2.2.3}).

To sample the posterior distribution $P(\boldsymbol{\omega}|\boldsymbol{d})$, we utilize the Markov Chain Monte Carlo (MCMC) method. Thereafter, we choose the location that maximizes the joint posterior probability as the best-fit value of the parameter. The sampling is performed using the Python package \texttt{emcee} (\citealt{Foreman-Mackey_2013_EMCEE}).

\subsection{Choices of Anisotropy}
As discussed in Sec.~\ref{sec:2.2.3}, Jeans modeling requires an assumption on stellar orbital anisotropy. For our mock data, we have adopted the logistic profile to fit the orbital anisotropy of the simulated galaxies and generate the mock central velocity dispersion. In the current analysis, we aim to investigate the biases arising from those commonly used anisotropy assumptions.

We choose three model scenarios with distinct anisotropy assumptions. We first adopt the isotropic orbit. In this case, the anisotropy parameter is fixed at 0, so no additional model parameter is introduced into $\boldsymbol{\omega}$. The second scenario is assuming the galaxy has a spatially uniform orbital anisotropy distribution, and the vector of the parameter space is extended to include the value anisotropy parameter $\beta^{\rm (mod)}$, becoming $\boldsymbol{\omega}=\{\log{M_\star/\rm M_\odot},f_{\rm dm},\eta_{\rm dm},\beta^{\rm (mod)}\}$. The third choice is using the parametric Osipkov-Merritt model, where the anisotropy varies as $\beta(r) = r^2/({r^2+r_{\rm om}^2})$. The transition radius $r_{\rm om}$ is re-parameterized in $\boldsymbol{\omega}$ as $\boldsymbol{\omega}=\{\log{M_\star/\rm M_\odot},f_{\rm dm},\eta_{\rm dm},\log{r_{\rm om}/R_{\rm eff}}\}$.

\section{Potential Bias from Orbit Anisotropy}
\label{sec:4}
\subsection{Isotropic Model}
\label{sec:4.1}
\begin{figure}
    \centering
    \includegraphics[width=\linewidth]{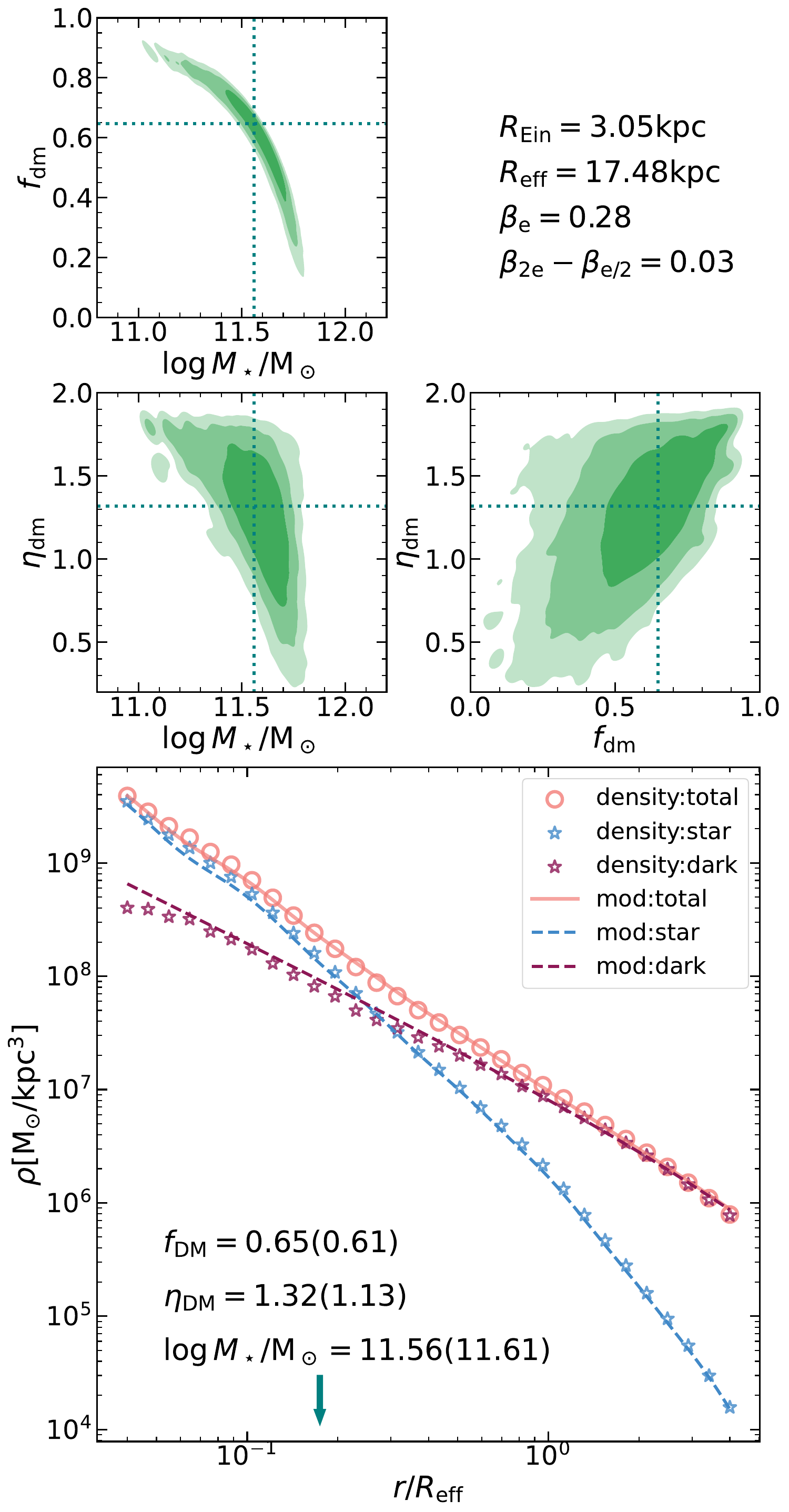}
    \caption{This is an example from the results of our joint lensing-dynamics modeling under the isotropic orbit assumption. This galaxy is drawn from the snapshot $z = 0.2$ (subfind ID: 230767). The three top panels present the combined posterior probability distributions where the contours represent $68.27\%$, $95.45\%$ and $99.73\%$ confidence levels. The dashed cross denotes the best-fit value of the parameter. And the bottom large panel shows the reconstruction for the density profile obtained from our best-fit model parameters. In the bottom panel, the pink, blue and purple points are the true total, stellar and dark matter density profile, respectively. The density profiles from our best-fit model are color-coded identically for comparison. We provide the inferred values of three model parameters followed by their true values (in bracket) directly derived from the simulation. The arrow indicates the position of the Einstein radius. We also list the Einstein radius, size, mock anisotropy as well as its gradient in the top-right corner.}
    \label{fig:example_iso}
\end{figure}

\begin{figure*}
    \centering
    \includegraphics[width=\textwidth]{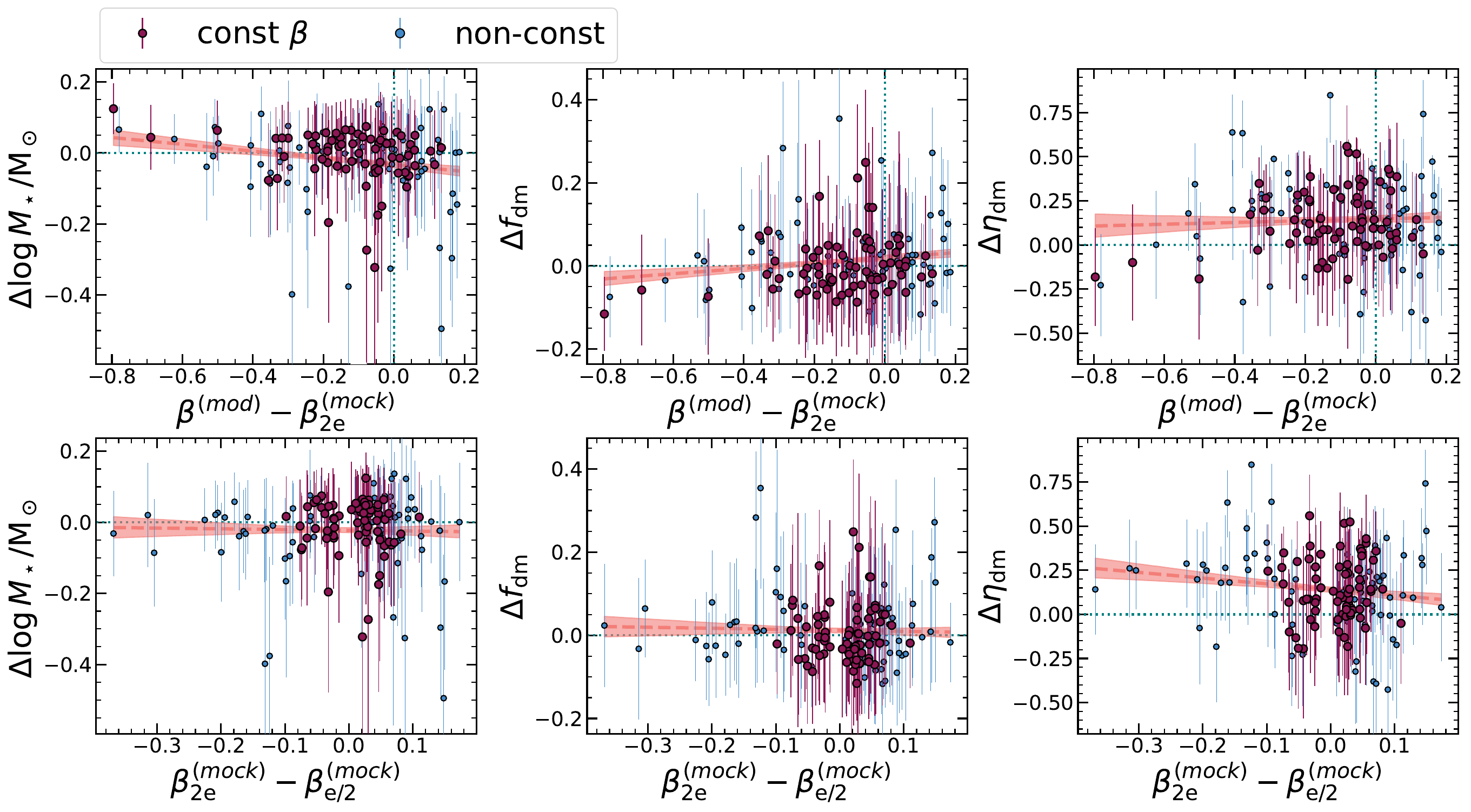}
    \caption{The statistical results of the lensing-dynamics modeling when we apply the anisotropy as a constant free parameter. Following the structure of Fig.~\ref{fig:results_iso}, we present the bias of the individual mass distribution parameter as the y-axis in the associated panel. And the x-axis of the bottom panel is still the anisotropy gradient of our mock galaxy. But the x-axis of the top panel represents the deviation between the model-predicted anisotropy parameter $\beta^{\rm (mod)}$ and the averaged anisotropy within $2R_{\rm eff}$. The purple and blue points are the galaxy with a constant anisotropy and other systems having a significant anisotropy gradient, respectively (see Sec.~\ref{sec:4.2}). The vertical dotted lines in the top panels denote the position of unbiased inference on the anisotropy itself.}
    \label{fig:results_const}
\end{figure*}

In the first scenario, we assume the galaxies have the isotropic stellar orbits (i.e., anisotropy $\beta^{\rm (mod)}=0$) in joint lensing-dynamics modeling, and the corresponding kernel function $K_\beta(u)$ in the Jeans equation solution (Eq~.\ref{eq:Jeans_solution}) is simplified to $K_\beta(u) = \sqrt{1-u^{-2}}$. The potential systematic biases may originate from two sources: the zeroth order is orbits deviating from the isotropic structure while the first order is introduced by the intrinsic gradient of the anisotropy parameter. Statistical results related to both aspects are shown in Fig.~\ref{fig:results_iso}. In the top panels, we present the systematic bias of each model parameter in vector $\boldsymbol{\omega}$ as a function of mock anisotropy averaged within the effective radius. And the bottom panels illustrate dependencies of biases on the anisotropy gradient. Here, the anisotropy gradient is still defined as the difference between the mock anisotropy parameter averaged within $2R_{\rm eff}$ and $0.5R_{\rm eff}$. The bias of the model parameter $\Delta\boldsymbol{\omega} = \{\Delta\log{M_\star/\rm M_\odot},\Delta f_{\rm dm},\Delta\eta_{\rm dm}\}$ is quantified as the discrepancy between the model-predicted value and the ground truth. The true values of $\log{M_\star/\rm M_\odot}$ and $f_{\rm dm}$ are obtained by directly summing the masses of the related particles. Specifically, the true value of the dark matter inner density slope is measured by fitting a gNFW model to the dark matter density profile within the virial radius ($r_{\rm dm}$ is free for this fitting). Generally speaking, while all model parameters exhibit scatter, the systematic biases at the population level are statistically insignificant. For clarity, we adopt the convention that a positive value of bias indicates an overestimation and a negative value of bias indicates an underestimation, with the associated uncertainty representing the intrinsic scatter of the population. The total stellar mass $\log M_\star/\rm M_\odot$ of the entire sample is slightly underestimated with $\Delta \log{M_\star/\rm M_\odot}=-0.028 \pm 0.118$ $\rm dex$ while systematic modeling bias in the central dark matter fraction $\Delta f_{\rm dm}$ remains $+1.4\% \pm 9.1\%$. The only quantity with a relatively more pronounced bias is the inner slope of the dark matter profile, which is steepened by $\Delta \eta_{\rm dm} = +0.18 \pm 0.22$, although consistent with no bias at $1 \sigma$ confidence level. 

\begin{figure}
    \centering
    \includegraphics[width=\linewidth]{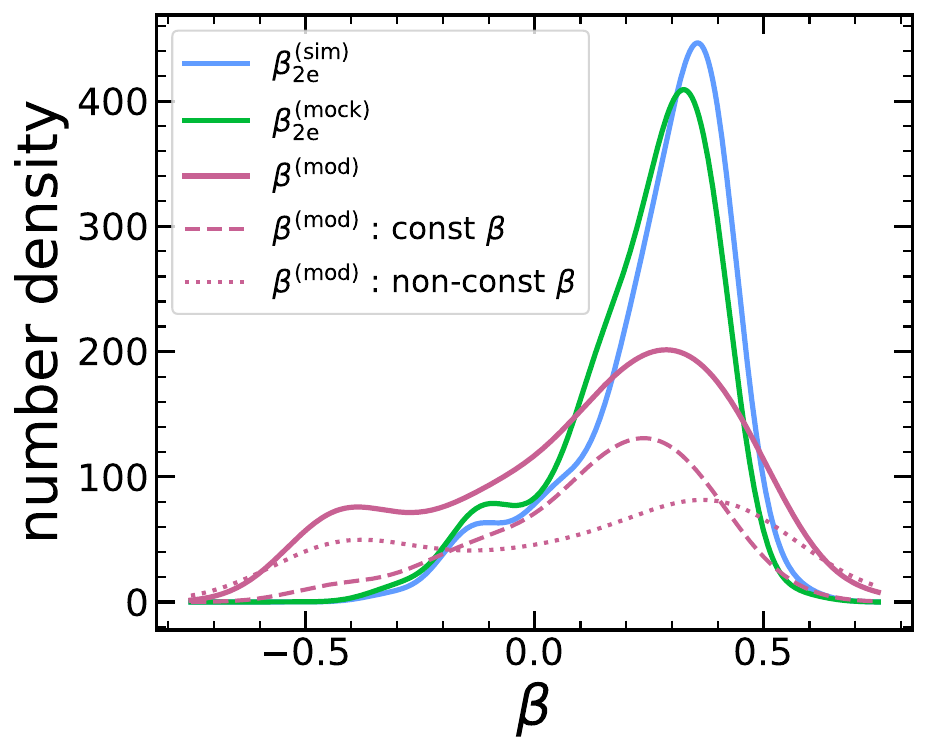}
    \caption{The smoothed number density distribution of stellar orbit anisotropy from the best-fit model compared to the averaged anisotropy of the mock systems. The number density distributions characterized by green and blue are the averaged mock anisotropy within $R_{\rm eff}$ and $2R_{\rm eff}$, respectively. The solid pink curve is the anisotropy parameter predicted by the joint lensing-dynamics model with a constant-$\beta$ anisotropy profile. We also present the modeling results of the galaxies with spatially constant anisotropy (dashed pink) and those with anisotropy profiles that exhibit significant radial variation (dotted pink).}   
    \label{fig:hist_const}
\end{figure}

\begin{figure*}
    \centering
    \includegraphics[width=\textwidth]{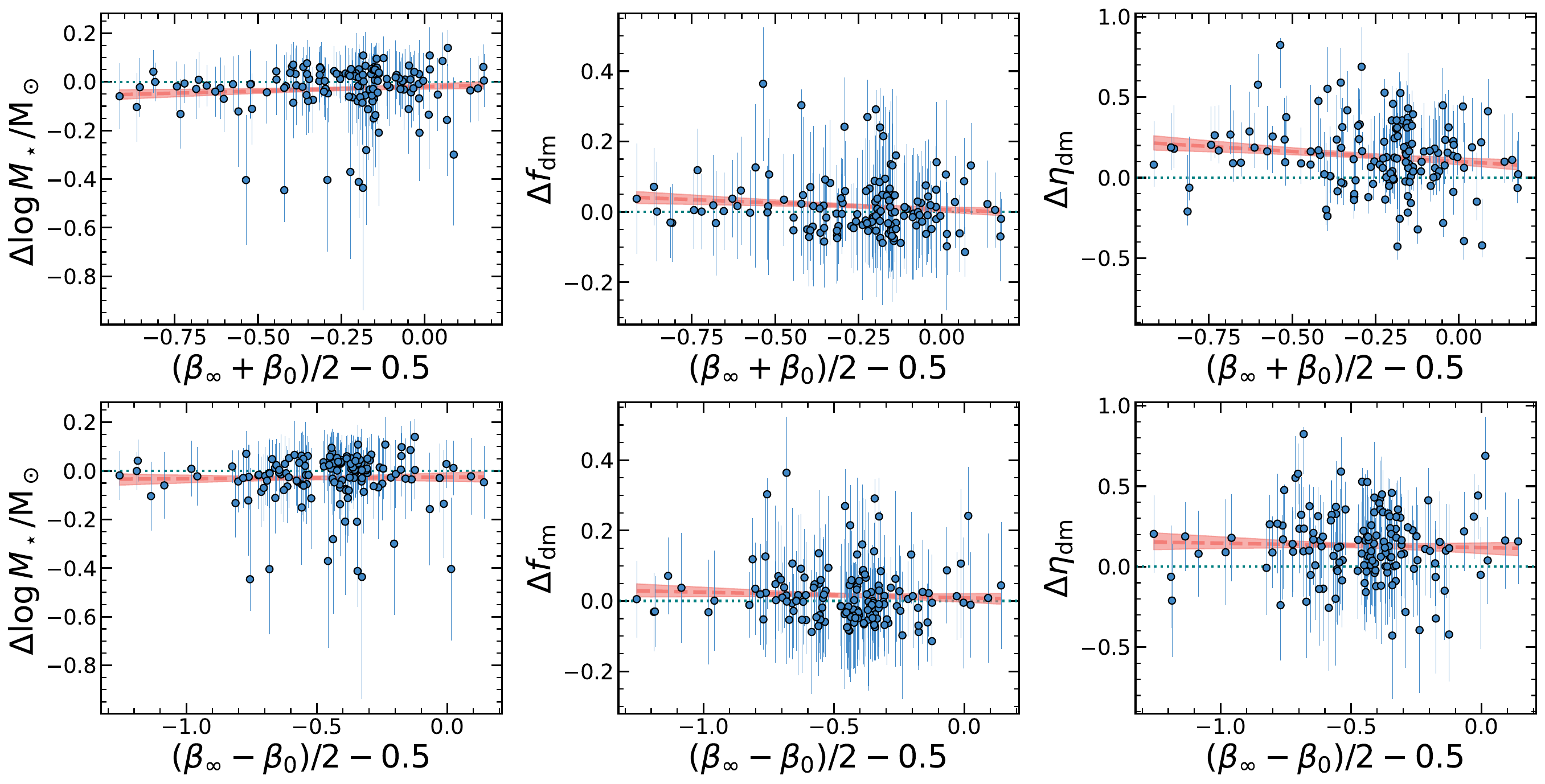}
    \caption{Modeling biases arising from the assumption of an Osipkov-Merritt anisotropy profile. The x-axes represent the deviations of the mock galaxy's logistic anisotropy profile from the Osipkov-Merritt model. All other plot specifications remain consistent with those presented in Fig.~\ref{fig:results_iso} and Fig.~\ref{fig:results_const}.}
    \label{fig:results_om}
\end{figure*}

\begin{figure}
    \centering
    \includegraphics[width=\linewidth]{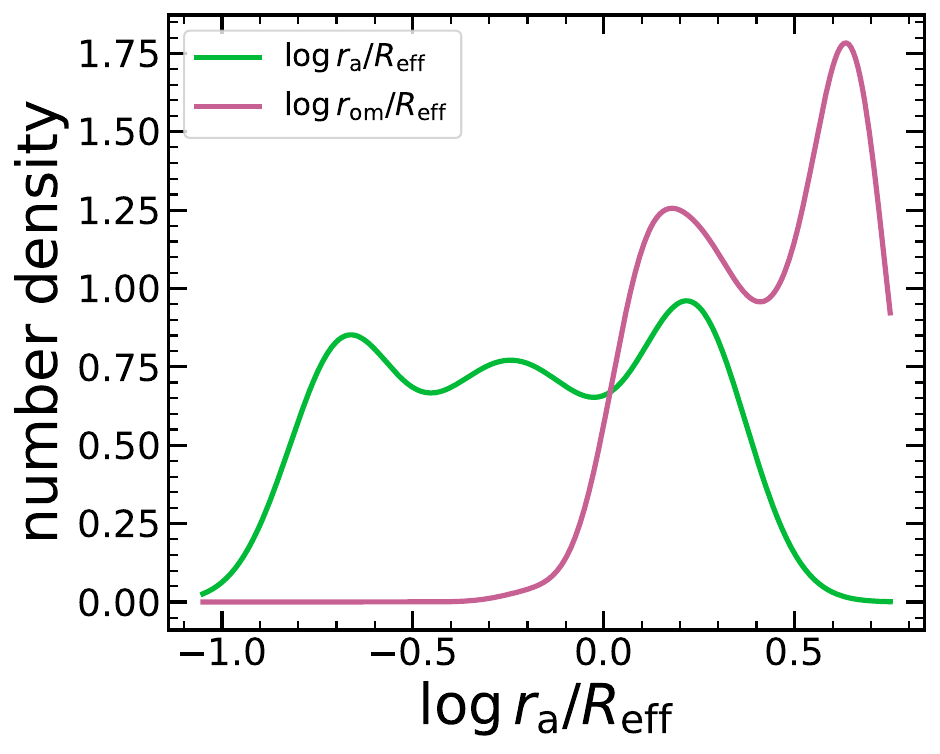}
    \caption{The distribution of the inferred value of the Osipkov-Merritt transition radius $r_{\rm om}$ (pink). For comparison, the number density illustrated by green is the transition radius of the logistic anisotropy profile during mocking processes.}
    \label{fig:hist_om}
\end{figure}

Additionally, to investigate the dependency of biases on orbital anisotropy, we highlight mock galaxies with an approximately isotropic orbital structure as the purple points in Fig.~\ref{fig:results_iso}, where a mock isotropic system is defined by two conditions: $|\beta_\infty-\beta_0|/2 < 0.15$ and $|\beta_\infty+\beta_0|/2 < 0.15$. The remaining galaxies are classified as anisotropic systems. Here, $\beta_0$ and $\beta_\infty$ are the inner and outer asymptotic limits of the mock logistic anisotropy profile (see Sec.~\ref{sec:2.2.3}). It can be seen that the anisotropic systems show a similar bias behavior with the full sample, where $\Delta\log M_\star/\rm M_\odot$ is $-0.032 \pm 0.122$ $\rm dex$ and the central dark matter fraction is overestimated with $\Delta f_{\rm dm} = + 1.8\% \pm 9.4\%$. The galaxies with isotropic orbits have $\Delta\log M_\star/\rm M_\odot = +0.016 \pm 0.043$ $\rm dex$ and $\Delta f_{\rm dm} = -2.5\% \pm 4.3\%$. Although the isotropic systems show opposing bias tendencies, the absolute magnitudes of the biases for both categories are relatively small. Furthermore, we present the linear regression results for the general trends between systematic biases in mass structure parameters and anisotropy discrepancy in Fig.~\ref{fig:results_iso}. This analysis also reveals no statistically significant correlation of systematic bias to the anisotropy parameter discrepancies. For instance, the linear slope $\partial(\Delta\log M_\star/\rm M_\odot)/\partial\beta_{\rm e}^{\rm (mock)}$ is $0.066$ with a corresponding p-value of 0.34 for the null hypothesis that the true slope equals to zero, indicating that $\Delta\log M_\star/\rm M_\odot$ is not correlated with $\beta_{\rm e}^{\rm (mock)}$. 

In Fig.~\ref{fig:example_iso}, we present a galaxy as an example. This galaxy is a typical massive ETG with a total stellar mass $\log{M_\star/\rm M_\odot} = 11.61$. Its stellar orbits is moderately radial dominated, characterized by $\beta^{(\rm mock)}_{\rm e} = 0.28$. With a biased inference at the average level, the true inner density slope of its dark matter halo is $\eta_{\rm dm} = 1.13$ while $\eta_{\rm dm}$ from the best-fit model is steepened to $1.32$ ($\Delta \eta_{\rm dm} = +0.19$). As a consequence, the central dark matter fraction is slightly overestimated by $4\%$ and the model under-predicts the stellar mass by 0.05 $\rm dex$. Despite these biases, the overall mass distribution is well recovered. We note that this galaxy is a representative of our sample, while galaxies with biases of a comparable or lower level can also be correctly recovered. Considering that the stellar masses measured using bottom-heavy IMF and bottom-light IMF in SPS modeling differ by almost a factor of two (\citealt{Smith_2020_M2L_IMF_variation}), this level of bias in stellar mass estimation does not significantly affect the assessment for the stellar IMF in those observational studies. 


\subsection{Constant-Anisotropy Model}
\label{sec:4.2}
The second scenario treats anisotropy as a constant free parameter throughout the modeling process. Under this assumption, predictions for the galaxy mass structure derived from lensing-dynamics modeling theoretically depend on both the anisotropy parameter inferred via posterior sampling and the intrinsic radial variation of the mock anisotropy profile. In this regard, we use Fig.~\ref{fig:results_const} to illustrate these potential dependencies: the top panels display modeling biases in mass distribution estimates as a function of the anisotropy bias itself, while bottom panels show the relationship between these biases and the anisotropy gradient. Here the bias of the anisotropy is defined as $\beta^{\rm (mod)}-\beta^{\rm (mock)}_{\rm 2e}$ and $\beta^{\rm (mod)}$ is the anisotropy predicted by the best-fit model. The systematic biases and scatters in this model exhibit no statistical changes compared to the isotropic case. Specifically, the stellar mass across the sample remains underestimated with $\Delta \log{M_\star/\rm M_\odot} = -0.023 \pm 0.100$ $\rm dex$ and the enhancement of the central dark matter fraction ($+1.2\% \pm 8.0\%$) is also subdominant (to its scatter). The inner slope of the dark matter density profile is also slightly biased, albeit insignificantly, toward steeper values, with $\Delta \eta_{\rm dm} = +0.14 \pm 0.22$. Moreover, as evidenced by the bottom panels of Fig.~\ref{fig:results_const}, there is no discernible dependence on the anisotropy gradient, e.g., the slope of the linear regression between $\Delta\log{M_\star/\rm M_\odot}$ and $\beta^{\rm (mock)}_{\rm 2e}-\beta^{\rm (mock)}_{\rm e/2}$ is $-0.021$ (p-value: $0.813$) while $\partial\Delta f_{\rm dm}/\partial(\beta^{\rm (mock)}_{\rm 2e}-\beta^{\rm (mock)}_{\rm e/2})$ is $-0.024$ with a p-value of $0.73$. This conclusion is further corroborated by the consistent fitting performances observed across systems with constant anisotropy (purple points, $|\beta_\infty-\beta_0|/2 < 0.15$) and those with non-constant anisotropy profiles (blue points,$|\beta_\infty-\beta_0|/2 > 0.15$). Both subsets have the averaged bias in stellar mass inference from $-0.010$ to $-0.035$ $\rm dex$ and $\Delta f_{\rm dm} \sim +2\%$.   

However, in this scenario, the inference of stellar orbit anisotropy itself proves to be inadequate. As illustrated in the top panels of Fig.~\ref{fig:results_const}, the overall anisotropy $\beta^{\rm (mod)}$ is significantly underestimated for a notable portion of galaxies. Fig.~\ref{fig:hist_const} presents a comparison between the anisotropy values derived from our model and those of the mock systems. Within the apertures of $2R_{\rm eff}$, the stellar kinematics of the most mock galaxies are dominated by the moderately radial orbits (green curve), consistent with the measurements from the simulation (blue curve). In contrast, the anisotropy predicted by the model displays a slightly bimodal distribution across the sample (purple curve). Some of the galaxies, particularly those featuring an anisotropy gradient (dotted curve), are fitted with a tangential value for the anisotropy. Our results imply that, when combining single-aperture velocity dispersion measurement with lensing observations, the radial variation in stellar orbit anisotropy does not significantly interfere with estimates for the stellar mass, dark matter fraction as well as its inner density slope. But the presence of an anisotropy gradient can introduce biases in the inference of anisotropy itself, as expected (since, if your model does not have the flexibility to reproduce the reality/ground truth, the outcome will be biased).

\subsection{Osipkov-Merritt Model}
In this model, we primarily consider the radial variation of the anisotropy via the Osipkov-Merritt model. As a special case of the logistic profile, the only free parameter is the transition radius. 
To evaluate the discrepancy between the Osipkov-Merritt model and the anisotropy profile of our mock sample, we employ two metrics: the anisotropy value at the transition radius, denoted as $(\beta_\infty+\beta_0)/2$, and half the width of the transition, expressed as $(\beta_\infty-\beta_0)/2$. These parameters collectively describe the variability of anisotropy. Notably, within the Osipkov-Merritt model, both metrics are 0.5. Therefore, we analyze the modeling biases in the stellar mass, dark matter fraction and the dark matter inner density slope as functions of the deviations $(\beta_\infty+\beta_0)/2-0.5$ and $(\beta_\infty-\beta_0)/2-0.5$ in Fig.~\ref{fig:results_om}. Our findings reveal that the modeling biases closely resemble those observed in the isotropic scenario. The overall bias of the total stellar mass is $-0.029 \pm 0.106$ $\rm dex$ while the central dark matter fraction is over-predicted by $1.6\% \pm 8.2\%$. Correspondingly, the model gives a steeper inner density slope with $\Delta \eta_{\rm dm} = +0.13\pm 0.21$. Furthermore, there is no significant correlation or discernible preference between the modeling biases and the extent to which the logistic mock anisotropy deviates from the Osipkov-Merritt model.

\begin{figure}
    \centering
    \includegraphics[width=\linewidth]{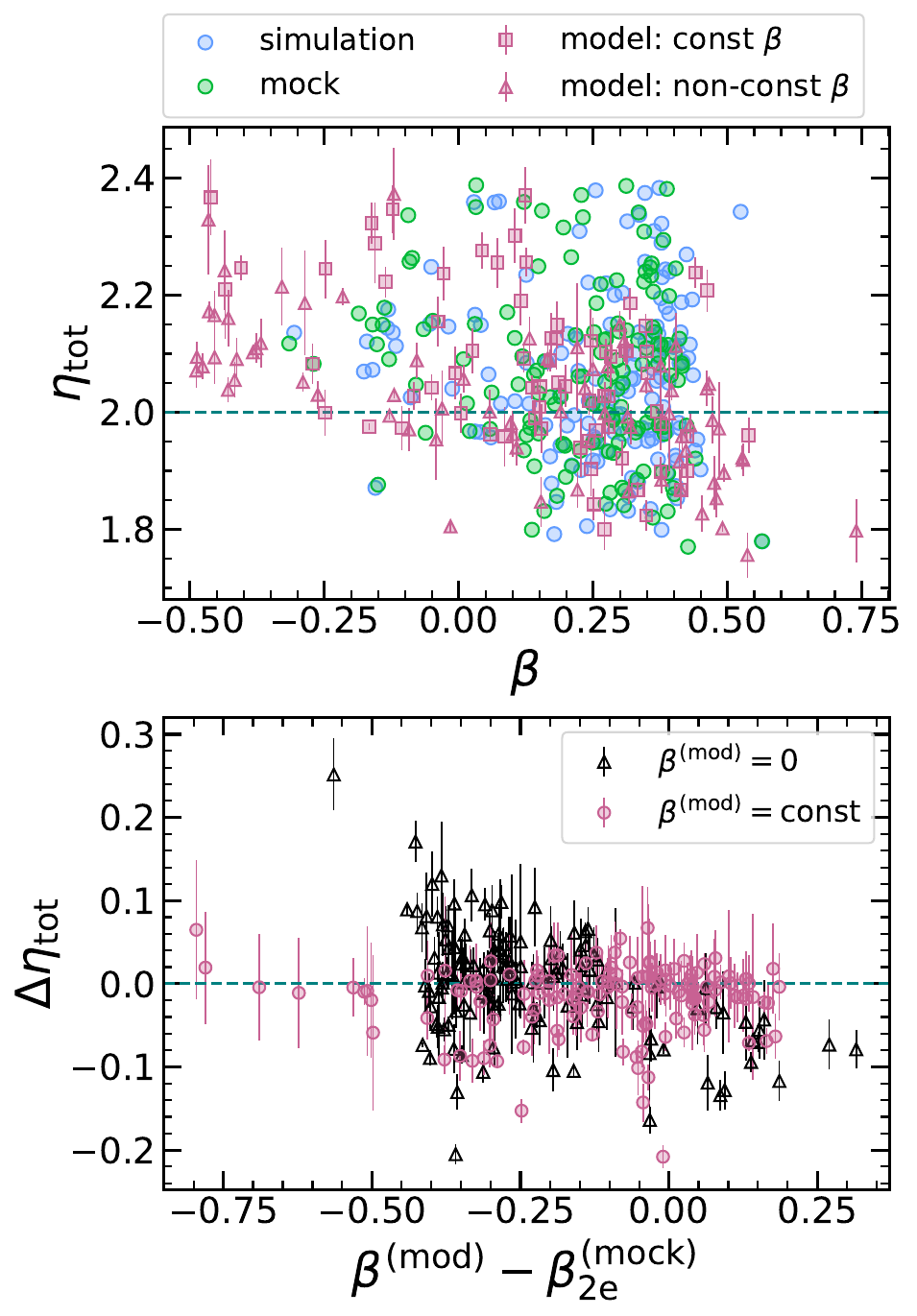}
    \caption{The logarithmic total density slope is presented as a function of stellar orbital anisotropy. In the top panel, blue points show direct measurements of both $\eta_{\rm tot}$ and $\beta_{\rm 2e}$ from simulations, while green points represent ground truth values from our mock sample. Purple data points illustrate the total density slope and anisotropy inferred from the best-fit results of our lensing-dynamics model, where anisotropy is treated as a free constant (see Sec.~\ref{sec:4.2}). Squares and triangles indicate whether the mock system exhibits approximately constant anisotropy. The bottom panel demonstrates how the bias in the total density slope $\Delta \eta_{\rm tot} = \eta^{\rm (mod)}_{\rm tot} - \eta_{\rm tot}^{\rm (mock)}$ correlates with the bias in anisotropy measurements $\beta^{\rm (mod)} - \beta_{\rm 2e}^{\rm (mock)}$. Black and purple points correspond to results from the isotropic scenario (Sec.~\ref{sec:4.1}, $\beta^{\rm (mod)} = 0$) and the constant anisotropy scenario (Sec.~\ref{sec:4.2}), respectively.}
    \label{fig:total_density_slope}
\end{figure}
We also present the inferred transition radius $r_{\rm om}$ from our model in Fig.~\ref{fig:hist_om}. For the mock galaxy sample, the anisotropy transition mainly occurs within the effective radius. In contrast, most of the model predictions have $r_{\rm om}$ exceeding $3R_{\rm eff}$. In other words, the inferred stellar motions are nearly isotropic in the region where the total mass profile is dominated by stars. Consequently, the lensing-dynamics model provides a similar prediction as the results under the isotropic scenario.

\section{Discussion}
\label{sec:5} 

\subsection{Dark Matter Inner Density Slope}
\label{sec:5.1}
Sec.~\ref{sec:4} shows that the dark matter inner density slope is overestimated by our model (by a factor of $0.13$-$0.18 \pm 0.2$) in all three scenarios. Moreover, as we have listed above, the bias in $\eta_{\rm dm}$ keeps the same order of magnitude under each anisotropy assumption. Besides, within each individual scenario, no dependency is found between the $\Delta \eta_{\rm dm}$ and the deviation from our mock anisotropy profile to the anisotropy assumptions applied in our lensing-dynamics model. For instance, $\partial\Delta\eta_{\rm dm}/\partial (\beta_{\rm 2e}^{\rm (mock)}-\beta_{\rm e/2}^{\rm (mock)})$ in isotropic and constant-anisotropic scenario are $0.15$ (p-value:$0.447$) and $-0.32$ (p-value: $0.09$), respectively. The observed bias in $\eta_{\rm dm}$ probably stems from other sources rather than assumptions on the anisotropy. 

\subsection{Comparison with Biases Arising from Stellar Mass-to-Light Ratio Gradient}
\label{sec:5.2}
The radial gradient of stellar mass-to-light ratio is another major source of the biases in the lensing-dynamics model. For comparison, the total stellar mass in the tests presented in this paper is slightly under-predicted, with $\Delta\log{M_\star/\rm M_\odot} \sim -0.03\pm0.1$ $\rm dex$. In contrast, biases in the total stellar mass (or mass-to-light ratio) arising from the stellar mass-to-light ratio gradient can reach up to $+0.2$ $\rm dex$ (\citealt{LiangYan_M2L_bias}). On the dark matter side, the central dark matter fraction is only overestimated by $3\% \pm 10\%$ whereas the bias in the dark matter fraction within $R_{\rm eff}$ caused by stellar mass-to-light ratio gradient is approximately $-20\%$ (\citealt{LiangYan_M2L_bias}). Our tests indicate that the biases resulting from the orbital anisotropy is less significant than the biases introduced by the stellar mass-to-light ratio gradient.

\subsection{Anisotropy versus Total Density Slope}
\label{sec:5.3}
In Sec.~\ref{sec:4.1}, we have shown the modeling bias under the isotropic orbit scenario. In Sec.~\ref{sec:4.2}, we have presented the results of assuming the anisotropy as a constant free parameter during modeling. Although the modeling biases for $\log M_\star/\rm M_\odot$ and $f_{\rm dm}$ are relatively modest, we observe a systematic underestimation of the anisotropy itself. Previous studies (\citealt{Koopmans_2009_SLACS_ETGs_Isotropy,XuDandan_2017_Illustris_ETGs}) have reported that adopting a constant but biased anisotropy can shift of the prediction on logarithmic total density slope $\eta_{\rm tot}$. Therefore, it is interesting to examine the logarithmic total density slope of our sample. 

Conventionally, the total density slope is defined as the power-law index, assuming a total density profile for the galaxy of the form $\rho_{\rm tot}(r) \propto r^{-\eta_{\rm tot}}$. However, the matter distribution of the simulated galaxy can be affected by the artificial core problem due to the force softening. To address this, we employ a cored power-law model to accurately measure the total density slope, as follows:
\begin{equation}
    \rho_{\rm tot}(r) = \frac{\rho_0 r_{\rm c}^{\eta_{\rm tot}}}{(r^2+r_{\rm c}^2)^{\eta_{\rm tot}/2}},
\end{equation}
where $r_{\rm c}$ is the core radius. We fit the cored power-law model to three density profiles: (1) the total density profile extracted directly from simulation particle data, (2) the mock density profile represented using the MGE model, and (3) the total density derived from our best-fit modeling (with stars modeled via MGE and dark matter via the gNFW profile). 

In Fig.~\ref{fig:total_density_slope}, we present the total density slope $\eta_{\rm tot}$ as a function of the anisotropy. The true values from simulation indicate no statistically significant correlation between $\eta_{\rm tot}$ and $\beta$, which is similar to the results for simulated ETGs at lower redshifts reported by \citet{WangYunchong_2020_ETG_TNG1}. In contrast, the $\eta_{\rm tot}$ values inferred from our lensing-dynamics model exhibit a weak anti-correlation with anisotropy, in agreement with the findings by \citet{Koopmans_2006_SLACS3_ETG_StructureFormation,Koopmans_2009_SLACS_ETGs_Isotropy}. We also investigate the dependency of bias in the total density slope $\Delta\eta_{\rm tot} = \eta_{\rm tot}^{\rm (mod)}-\eta_{\rm tot}^{\rm (mock)}$ on bias in the anisotropy $\beta^{\rm (mod)}-\beta^{\rm (mock)}_{\rm 2e}$. In the bottom panel of Fig.~\ref{fig:total_density_slope}, the bias of the total density slope exhibits no correlation with $\beta^{\rm (mod)}-\beta_{\rm 2e}^{\rm (mock)}$ if anisotropy is a constant free parameter (pink circles). The average bias in total density slope is merely $-0.02\pm 0.04$. This robust performance stems from two key factors: (1) our elimination of stellar mass-to-light ratio gradient interference and (2) the additional constraining power provided by the lensing observable  $\xi = \theta_{\rm E}\alpha^{\prime\prime}(\theta_{\rm E})/(1-\kappa(\theta_{\rm E}))$. If we assume $\beta^{\rm (mod)} = 0$ (black triangles), the average $\Delta\eta_{\rm tot}$ is $0.00\pm0.06$. But the isotropic assumption leads to a systematic underestimation on $\eta_{\rm tot}$ for the subsample with intrinsically tangential orbits ($\beta_{\rm 2e}^{\rm (mock)} < 0$) as also noted by \citet{XuDandan_2017_Illustris_ETGs}.

\section{Conclusions}
\label{sec:6}
In this study, we investigate possible systematic biases that may emerge in joint lensing-dynamics modeling due to assumptions made about stellar orbital anisotropy. We utilize the simulated massive ETGs from TNG100 simulation (\citealt{Nelson_TNG_1,Marinacci_TNG_2,Naiman_TNG_3,Pillepich_TNG_4,Springel_TNG_5}) as our galaxy sample, generating the mock lensing and stellar kinematics observables. To constrain our model, we combine strong lensing observables with the central stellar velocity dispersion and weak lensing signals. In order to accurately represent the stellar orbital anisotropy of the simulated galaxy sample, we employ the logistic profile (\citealt{Baes&vanHese_2007_LogisticAnisotropy}) to depict the radial distribution of the stellar orbital anisotropy. In our joint lensing-dynamics model, we adopt the MGE and the gNFW profile to describe the stars and dark matter components. The shape of the MGE for the stellar component is scaled according to the true stellar mass density distribution to mitigate interference from the stellar mass-to-light ratio gradient and thus isolate the impact of anisotropy. Overall, we implement three commonly-used assumptions regarding anisotropy in our analysis: isotropic orbits, constant anisotropy and the Osipkov-Merritt model. Our primary findings are summarized as follows:  

\begin{itemize}
    \item Across all three tested anisotropy assumptions, the total stellar mass $\log M_\star/\rm M_\odot$ shows a systematic bias of approximately $0.02$-$0.03$ $\pm 0.12 \rm dex$. Correspondingly, the central dark matter fraction is on average overestimated by $1.2\%$-$1.8\% \pm 10\%$. Notably, these biases in decomposing stellar and dark matter masses inside the halo center are significantly smaller than those introduced by the mass-to-light ratio gradient (\citealt{LiangYan_M2L_bias}).

    \item The lensing-dynamics model under all three anisotropy assumptions would predict a more centrally concentrated dark matter halo with a logarithmic inner density slope $\eta_{\rm dm}$ steepened by $0.13$-$0.18$ with a scatter of $0.2$ relative to the simulation.  

    \item No strong evidence suggests that the bias behavior depends on the discrepancy in the anisotropy assumptions and true orbital anisotropy of the mock systems.

    \item Modeling the orbital anisotropy as a free constant parameter fails to recover the correct average anisotropy value in galaxies which exhibit radial variations in their anisotropy profile. 
\end{itemize}

We note that compared to \citealt{LiangYan_M2L_bias}, the biases on the inferred stellar and dark matter mass normalizations and inner slopes due to the assumptions on stellar orbit anisotropy are significantly weaker than those due to the assumption of a constant stellar mass-to-light ratio. This is achieved in our tests by directly taking the stellar mass profile as the luminosity profile, i.e., fixing a constant stellar mass-to-light ratio, which helps to significantly reduce the systematic bias. The two studies combined suggest that the precision of stellar-dark matter decomposition is largely determined by the accuracy of the stellar mass-to-light ratio gradient. Therefore, future studies should aim to obtain more stringent constraints on the spatial distribution of the stellar mass-to-light ratio, while the assumption on the stellar orbit anisotropy is only a secondary effect.

The limitations of this work come from two sources. The first is the stellar orbits within the simulated galaxies. As discussed in Sec.~\ref{sec:2.2.1}, numerical effects, such as force softening due to resolution constraints and energy redistribution between stellar and dark matter particles (\citealt{Ludlow_2019_EnergyRepartition_Simulation,Ludlow_2023_SpuriousHeating_Simulation}). In our work, we observe that stellar orbital anisotropy tends to converge toward an isotropic state in the innermost regions ($<0.1R_{\rm eff}$). Undoubtedly, this structural characteristic shall leave an imprint in our mock kinematics. However, whether this isotropic core is real or not remains an open question. It is worthwhile to examine the stellar orbital anisotropy profile across the simulations of different resolutions in the upcoming work to clarify this issue. The second limitation concerns the accurate representation of orbital anisotropy in mock kinematics. While the logistic function for anisotropy parameterization has enhanced flexibility in the Jeans modeling framework, it remains incapable of capturing non-monotonic features in the anisotropy profile. To address this, one could potentially allow each stellar multi-Gaussian component to possess an individual anisotropy parameter. We leave this to a future study.  

\begin{acknowledgments}
We thank Dr. Kai Zhu for his helpful discussions and constructive suggestions. This work is supported by the National Natural Science Foundation of China (Grant No. 12433003) and the China Manned Space Project (No. CMS-CSST-2021-A07 and No. CMS-CSST-2025-A10). This work acknowledges the Tsinghua Astrophysics High-Performance Computing platform for providing the computational and storage resources that supported this research. 
\end{acknowledgments}

\software{astropy \citep{2013A&A...558A..33A,2018AJ....156..123A,2022ApJ...935..167A},  
          jampy \citep{Cappellari_2008_JAM_cylinderical,Cappellari_2020_JAM_spherical}, 
          mgefit \citep{Cappellari_2002_MGE},
          emcee \citep{Foreman-Mackey_2013_EMCEE}
          }


\appendix

\section{The Mock Anisotropy Via Logistic Profile}
\label{sec:mock_anisotropy}
We construct the mock stellar kinematics data from the simulated galaxies to match two quantities: (1) the line-of-sight velocity dispersion profile within $R_{\rm eff}$ (sampled with 10 linearly spaced radial bins), and (2) the stellar anisotropy profile within $2R_{\rm eff}$ (20 linearly spaced bins). These data are fitted with the spherical Jeans model (Eq.~\ref{eq:Jeans_solution}) and the logistic function for anisotropy (Eq.~\ref{eq:logistic}). The free parameters are $\{\beta_0,\beta_\infty,r_{\rm a}, a\}$ in Eq.~\ref{eq:logistic}. The optimization is performed by minimizing the composite $\chi^2$ term:
\begin{equation}
\label{eq:chi_mock_kinematics}
    \chi^2 = \sum\frac{(\sigma_{\rm i}^{\rm (mock)}-\sigma_{\rm i}^{\rm (sim)})^2}{\epsilon_{\sigma,i}^2}+\sum\frac{(\beta_{\rm i}^{\rm (mock)}-\beta_{\rm i}^{\rm (sim)})^2}{\epsilon_{\beta,i}^2}.
\end{equation}
where $\sigma_{\rm i}^{\rm (sim)}$ and $\beta_{\rm i}^{\rm (sim)}$ denote the line-of-sight velocity dispersion and anisotropy parameters derived from stellar particle motions in the corresponding radial bin while $\sigma_{\rm i}^{\rm (mock)}$ and $\beta_{\rm i}^{\rm (mock)}$ represent the corresponding values from the solution of Jeans equation and logistic anisotropy model. The mock velocity dispersion relies on the mock density distribution and the logistic anisotropy, the former have been obtained via the MGE fitting while the latter is also constrained by $\beta^{\rm (sim)}$. The uncertainties of both two quantities are determined through resampling of stellar particles within each radial bin. 

\begin{figure*}
    \centering
    \includegraphics[width=\textwidth]{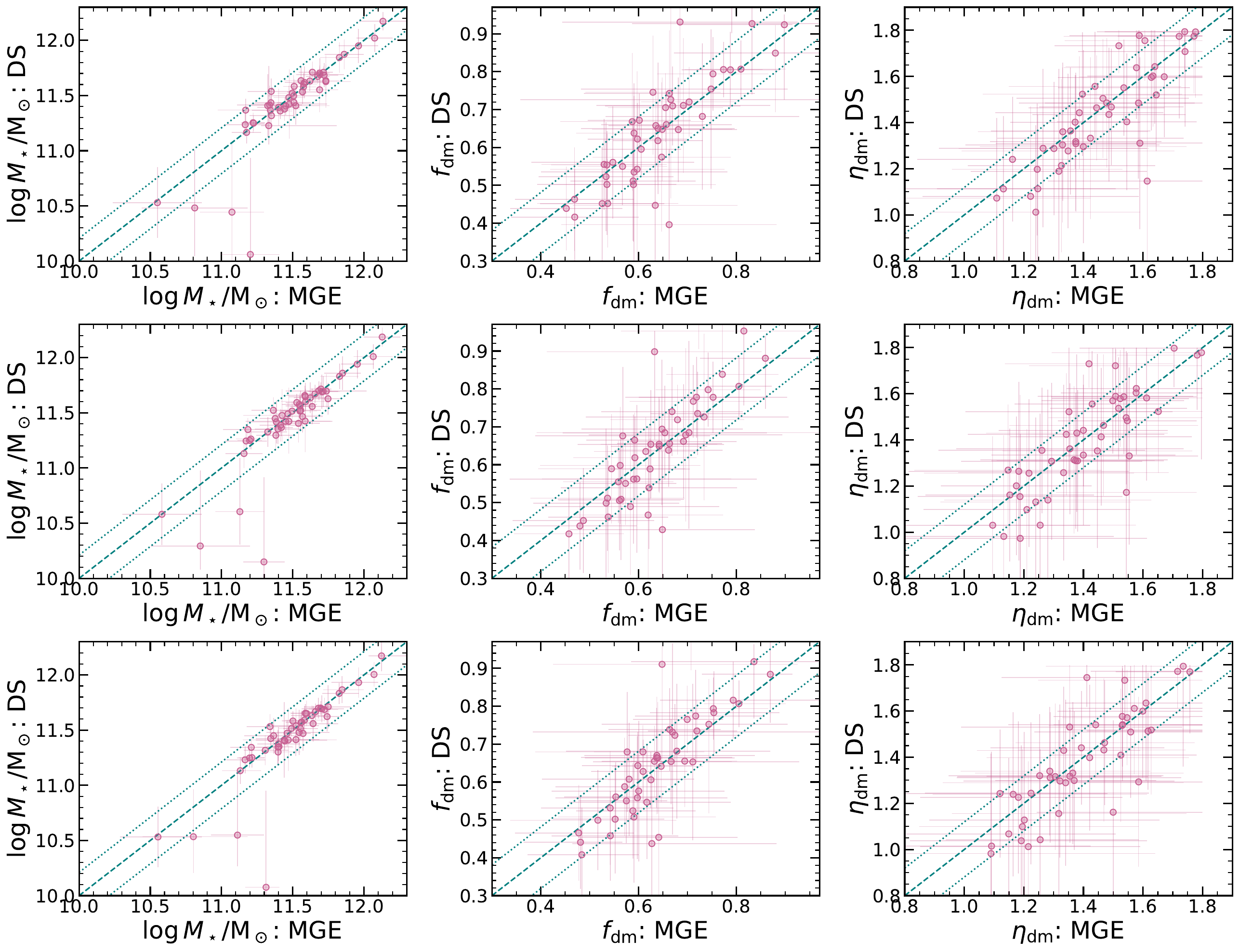}
    \caption{The comparison between using the parametric model (double-S{\'e}rsic profile, y-axis) and using the MGE (x-axis). \textit{Top:} isotropic scenario, \textit{Mid:} anisotropy treated as a free parameter and \textit{Bottom:} Osipkov-Merritt model. The dashed line denotes $y=x$ while the dotted lines indicate $1\sigma$ scatter of the data relative to the $y=x$.}
    \label{fig:comparison_mge_para}
\end{figure*}

In addition, we take the anisotropy averaged within some specified radius as our evaluation metric. The averaged anisotropy from simulations can be easily computed. For example, to estimate the value of $\beta_{\rm e}^{\rm (sim)}$, we simply aggregate all stellar particles within the effective radius $R_{\rm eff}$ and calculate the mass-weighted second moment of their radial and tangential velocities. For the averaged mock anisotropy $\beta_{\rm e}^{\rm (mock)}$, once we have determined the optimal parameters of the logistic anisotropy model by minimizing Eq.~\ref{eq:chi_mock_kinematics}, we proceed to calculate the mass-weighted second moment of the radial velocity within $R_{\rm eff}$ as follows:
\begin{equation}
    \langle \overline{v^2_{\rm r}}(<R_{\rm eff})\rangle = \frac{\int_{0}^{R_{\rm eff}}4\pi r^2 \rho_{\rm st}^{\rm (\rm mock)}(r)\overline{v^2_{\rm r}}(r){\rm d}r}{\int_{0}^{R_{\rm eff}}4\pi r^2 \rho_{\rm st}^{\rm (\rm mock)}(r){\rm d}r},
\end{equation}
where $\overline{v^2_{\rm r}}(r)$ is derived from the Jeans model with the best-fit parameters characterizing logistic anisotropy model. Similarly, according to the definition of orbital anisotropy, the tangential velocity dispersion averaged within $R_{\rm eff}$ is given by 
\begin{equation}
    \langle \overline{v^2_{\rm t}}(<R_{\rm eff})\rangle = \frac{\int_{0}^{R_{\rm eff}}4\pi r^2 \rho_{\rm st}^{\rm (\rm mock)}(r)\overline{v^2_{\rm r}}(r)(1-\beta(r)){\rm d}r}{\int_{0}^{R_{\rm eff}}4\pi r^2 \rho_{\rm st}^{\rm (\rm mock)}(r){\rm d}r},
\end{equation}
where $\beta(r)$ is the logistic model. Consequently, $\beta_{\rm e}^{\rm (mock)}$ can be expressed as:
\begin{equation}
    \beta_{\rm e}^{\rm (mock)} = 1-\frac{\langle \overline{v^2_{\rm t}}(<R_{\rm eff})\rangle}{\langle \overline{v^2_{\rm r}}(<R_{\rm eff})\rangle}.
\end{equation}
This implies that $\beta_{\rm e}^{\rm (mock)}$ is actually weighted by the factor $4\pi r^2 \rho_{\rm st}^{\rm (\rm mock)}(r)\overline{v^2_{\rm r}}(r)$, leading to the following formulation:
\begin{equation}
    \beta_{\rm e}^{\rm (mock)}= \frac{\int_{0}^{R_{\rm eff}}4\pi r^2 \rho_{\rm st}^{\rm (\rm mock)}(r)\overline{v^2_{\rm r}}(r)\beta(r){\rm d}r}{\int_{0}^{R_{\rm eff}}4\pi r^2 \rho_{\rm st}^{\rm (\rm mock)}(r)\overline{v^2_{\rm r}}(r){\rm d}r}.
\end{equation}

\section{Double S\'ersic Model For Stars}
\label{sec:double_sersic}

In this study, we employ the multi-Gaussian expansion (MGE) approach to represent the stellar component within our modeling framework, simulating the constraints from stellar light. However, non-parametric models, such as the multi-Gaussian model, need photometric data with a relatively high spatial resolution. Consequently, in numerous prior studies, parametric models, including the Hernquist model (\citealt{Hernquist_1990_HernProfile}), de Vaucouleurs model (\citealt{deVaucouleurs_1948}), and S{\'e}rsic profile (\citealt{Sersic_1963}), have also been utilized to characterize stellar light. To assess the robustness of our findings, we further conduct experiment with substituting the MGE model with the double-S{\'e}rsic model. We also apply the same sample selection criteria (as detailed in Sec.~\ref{sec:2.2.1}) based on fitting quality and projection performance when we fit the stellar mass density profile with the double-S{\'e}rsic model. Only the galaxies well described by both double-S{\'e}rsic and MGE model are retained for this analysis. The results are presented in Fig.~\ref{fig:comparison_mge_para}. It is evident that the predictions under using the parametric model are consistent with the results obtained from the MGE model. 


\bibliography{sample701}{}
\bibliographystyle{aasjournalv7}



\end{document}